\documentclass[lettersize,journal]{IEEEtran}
\usepackage{amsmath}
\usepackage{amssymb}
\usepackage{times}
\usepackage{graphicx}
\usepackage[colorlinks=true, allcolors=blue]{hyperref}
\usepackage{booktabs}
\usepackage{array}
\usepackage{makecell}
\usepackage{pifont}
\usepackage{caption}
\usepackage{authblk}
\usepackage[T1]{fontenc}
\usepackage{multirow}
\usepackage{hyperref}
\usepackage{longtable}    
\usepackage{tabularx}
\usepackage{booktabs}
\usepackage{pifont}   
\usepackage{array}    
\usepackage{comment}
\usepackage{subcaption} 
\usepackage{microtype}  
\usepackage[table]{xcolor}
\usepackage{tabularx}
\usepackage{booktabs}

\usepackage{booktabs}      
\usepackage{tabularx}      
\usepackage{ragged2e}      
\usepackage{makecell}      
\usepackage{caption}       

\usepackage{tikz}

\UseRawInputEncoding

\usepackage{makecell}
\setcellgapes{4pt}
\makegapedcells

\usepackage[english]{babel}

\AtBeginDocument{%
  \fontdimen2\font=0.21em 
  \fontdimen3\font=0.1em  
  \fontdimen4\font=0.06em  
}

\begin{document}
\bstctlcite{IEEEexample:BSTcontrol}
\title{Industrial Data-Service-Knowledge Governance: Toward Integrated and Trusted Intelligence} 

\author{
        Hailiang~Zhao,~\IEEEmembership{Member,~IEEE,} 
        Ziqi~Wang,
        Daojiang~Hu,~Mingyi~Liu,~Jiahui~Zhai,~Kai~Di,~Xinkui~Zhao,
        Zhongjie~Wang, 
        Jianwei~Yin,
        Albert~Zomaya,~\IEEEmembership{Fellow,~IEEE},
        MengChu~Zhou,~\IEEEmembership{Fellow,~IEEE},
        Shuiguang~Deng,~\IEEEmembership{Senior~Member,~IEEE}
\thanks{This work was supported by the National Key Research and Development Program of China under Grant 2024YFB3309400.}
\thanks{Hailiang Zhao, Ziqi Wang, Daojiang Hu, Xinkui Zhao, and Jianwei Yin are with the School of Software Technology, Zhejiang University. Emails: \{hliangzhao, wangziqi0312, daojianghu, zhaoxinkui, zjuyjw\}@zju.edu.cn.}
\thanks{Mingyi Liu and Zhongjie Wang are with the Faculty of Computing, Harbin Institute of Technology. Emails: \{liumy, rainy\}@hit.edu.cn.}
\thanks{Jiahui Zhai is with the College of Computer Science, Beijing University of Technology. Emails: zhaijiahui@emails.bjut.edu.cn.}
\thanks{Kai Di is with the Hangzhou School of Automation, Zhejiang Normal University. Email: dikai1994@zjnu.edu.cn.}
\thanks{Albert Y. Zomaya is with the School of Computer Science, The University of Sydney, Sydney, NSW 2006, Australia. Email: albert.zomaya@sydney.edu.au.}
\thanks{M. Zhou is with the Department of Electrical and Computer Engineering, New Jersey Institute of Technology, Newark, NJ 07102, USA. Email: zhou@njit.edu.}
\thanks{Shuiguang Deng is with the College of Computer Science and Technology, Zhejiang University. Email: dengsg@zju.edu.cn.}
}



\maketitle

\begin{abstract}

The convergence of artificial intelligence, cyber-physical systems, and distributed networking has accelerated the evolution of industrial intelligence across edge, cloud, and cross-organizational communication environments. However, existing governance mechanisms remain fragmented across data management, service orchestration, and knowledge-based decision-making, making it difficult to ensure reliability, accountability, compliance, and explainability throughout the industrial intelligence stack. To address this gap, we present \textsc{TRISK} (TRusted Industrial Data-Service-Knowledge governance), a conceptual and taxonomic framework for trustworthy industrial intelligence. \textsc{TRISK} is grounded in a five-dimensional trust model covering quality, security, privacy, fairness, and explainability, and formalizes how trust is constructed, propagated, aggregated, and fed back across data, service, and knowledge layers in networked industrial systems. Through a structured synthesis of more than 100 representative studies, standards, 
and technical reports, we examine data governance as the foundation of trust construction, service governance as the mediation layer for trustworthy execution, and knowledge governance as the semantic anchor for reasoning, validation, and feedback adaptation. We further discuss industrial implementation patterns, cross-industry implications, and the role of emerging communication and computing technologies. Finally, we outline a future research roadmap toward adaptive, verifiable, and human-aligned industrial governance for Industry 5.0.

\end{abstract}


\section{Introduction} \label{sec:intro}

\begin{figure}[!t]
    \centering
    \includegraphics[width=0.6\linewidth]{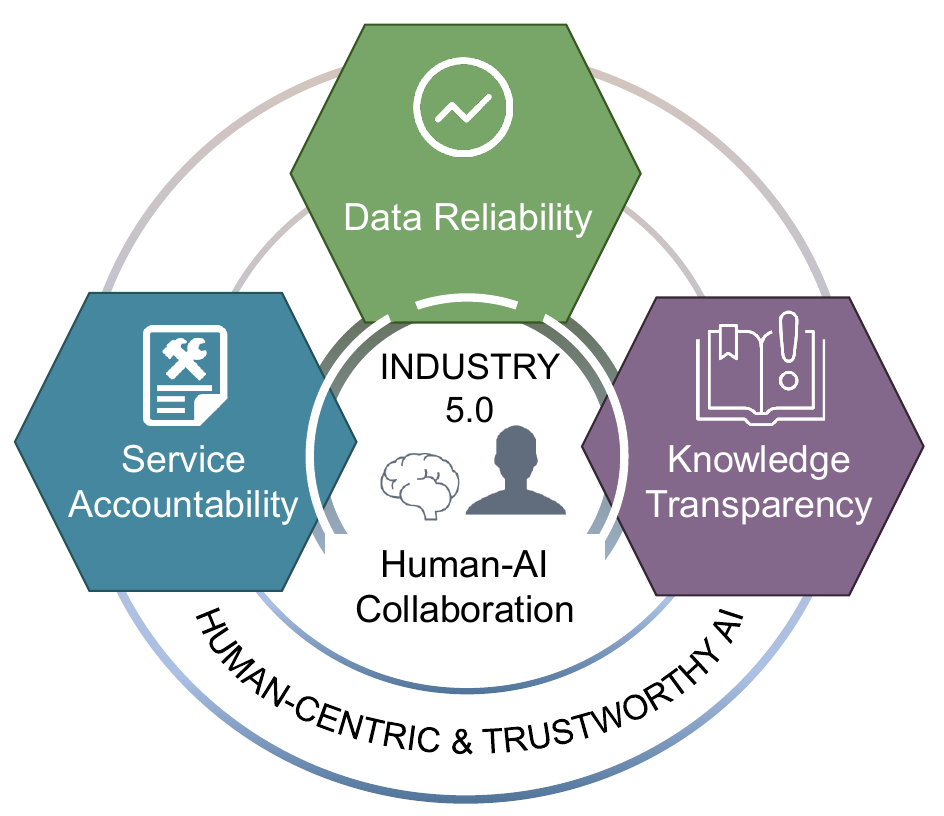}
    \caption{Trustworthiness dimensions in human-centric industrial intelligence under the Industry~5.0 paradigm. Trust is not isolated but propagates across data, network, service, and knowledge layers, constrained by communication dynamics and physical safety requirements.}
    \label{intro}
\end{figure}

The rapid digitalization of industrial ecosystems, driven by Industry 4.0 and the human-centric vision of Industry 5.0, has fundamentally transformed cyber-physical production systems into highly distributed, networked intelligence fabrics. Modern manufacturing, energy, and transportation infrastructures now rely on the tight convergence of large-scale data acquisition~\cite{Tao2022DigitalTwinReview}, cloud-edge collaborative computing~\cite{smcziqi,zhao2025online}, and intelligent service orchestration over deterministic communication networks~\cite{Lee2023IndustrialAI}. While this integration enables unprecedented autonomy, it introduces deeply intertwined trustworthiness challenges that span the entire stack, from physical sensing and network transmission to service execution and knowledge reasoning~\cite{Chen2023TrustworthyIIoT,zhang2024exploiting} (Fig.~\ref{intro}). Critically, unlike purely digital systems, industrial intelligence operates under irreversible physical consequences and strict real-time constraints. Trust can no longer be treated as a post-hoc audit property; it must be verifiably embedded and continuously propagated across heterogeneous networked layers, where communication latency, jitter, and protocol heterogeneity directly impact data fidelity, service accountability, and knowledge transparency.

\begin{figure*}[!t]
  \centering
  \begin{subfigure}[b]{0.35\textwidth}
    \centering
    \includegraphics[width=\linewidth]{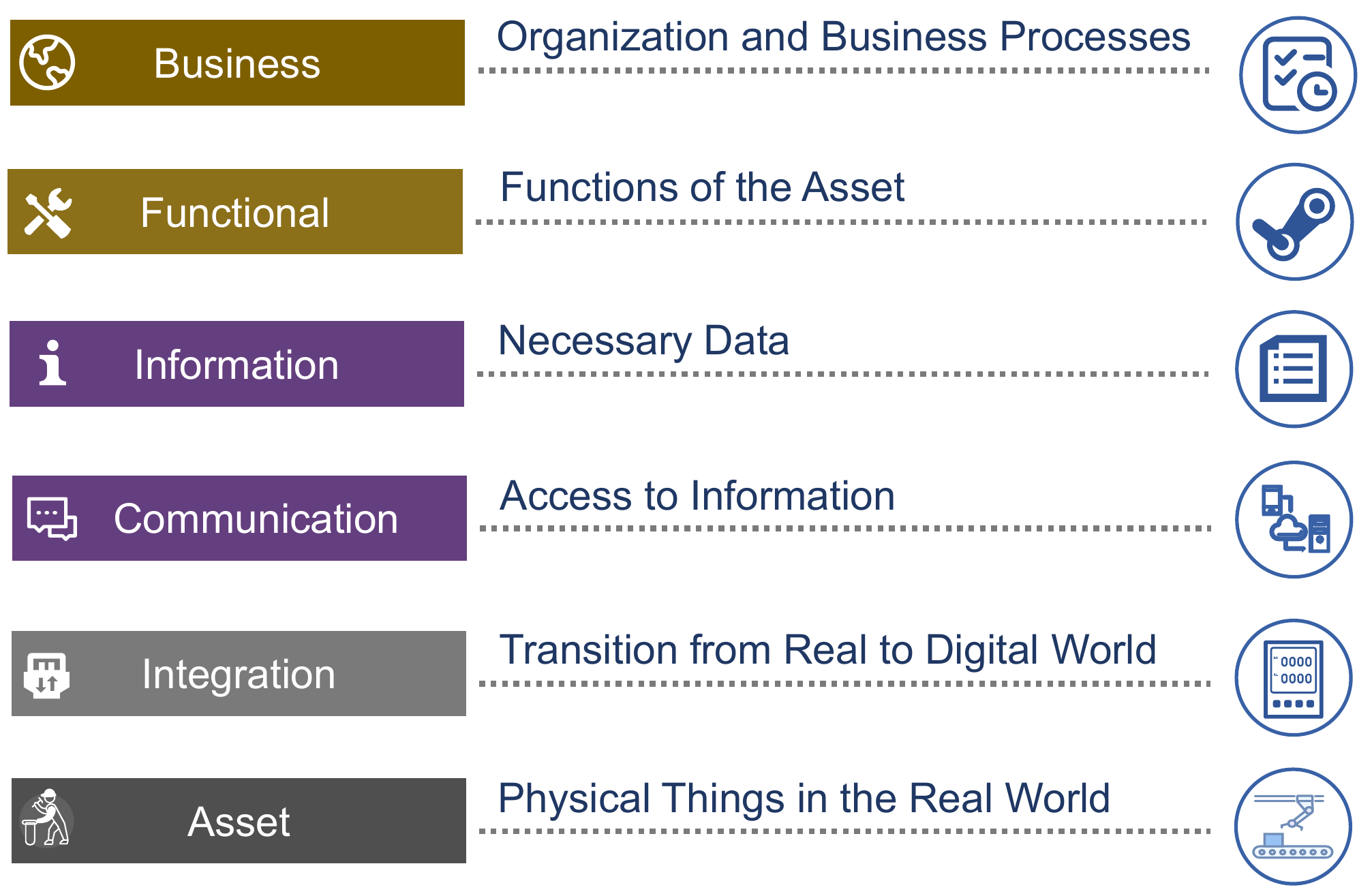}
    \caption{RAMI 4.0}
    \label{fig:rami}
  \end{subfigure}\hfill
  \begin{subfigure}[b]{0.33\textwidth}
    \centering
    \includegraphics[width=\linewidth]{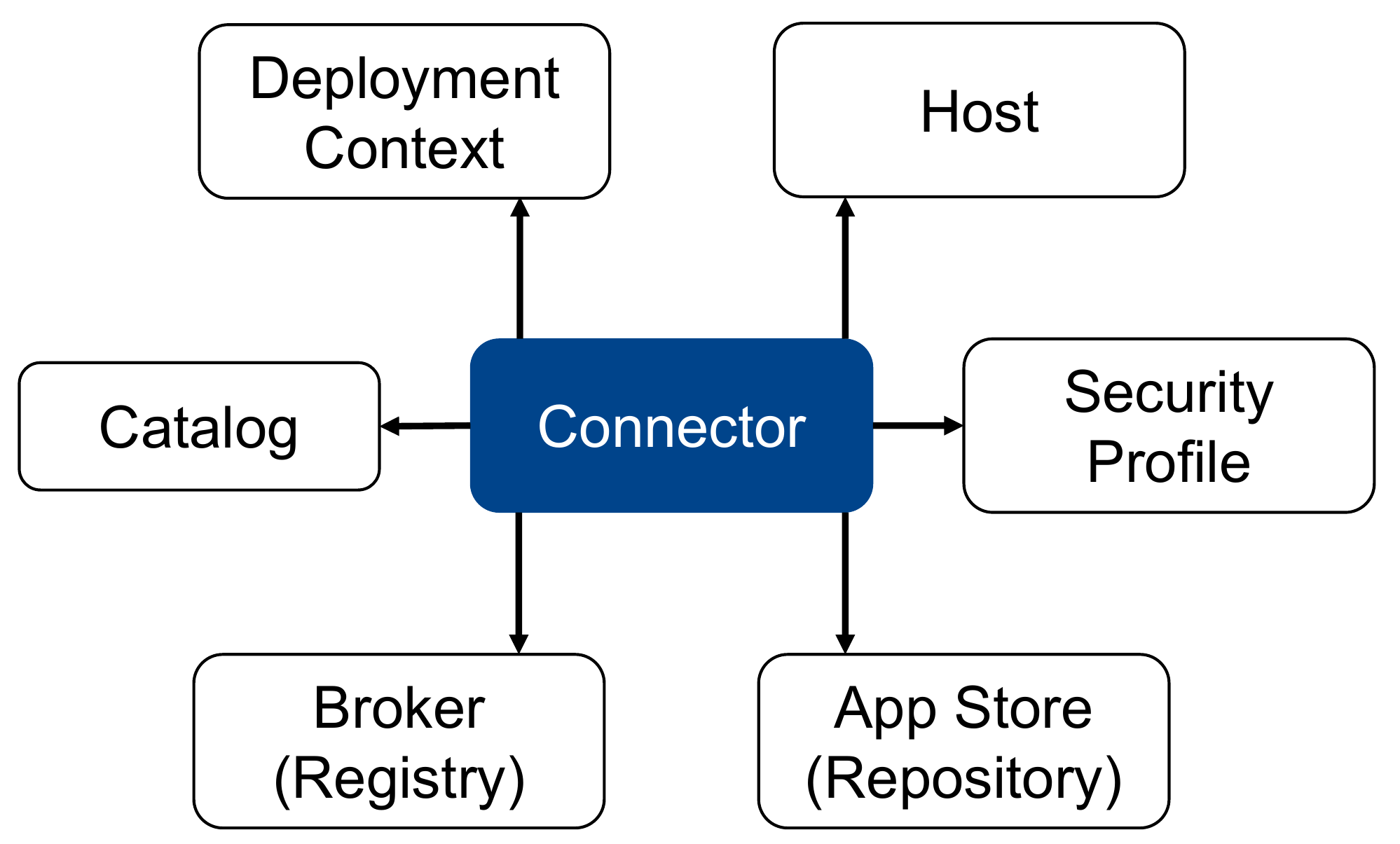}
    \caption{IDS}
    \label{fig:ids}
  \end{subfigure}\hfill
  \begin{subfigure}[b]{0.3\textwidth}
    \centering
    \includegraphics[width=0.94\linewidth]{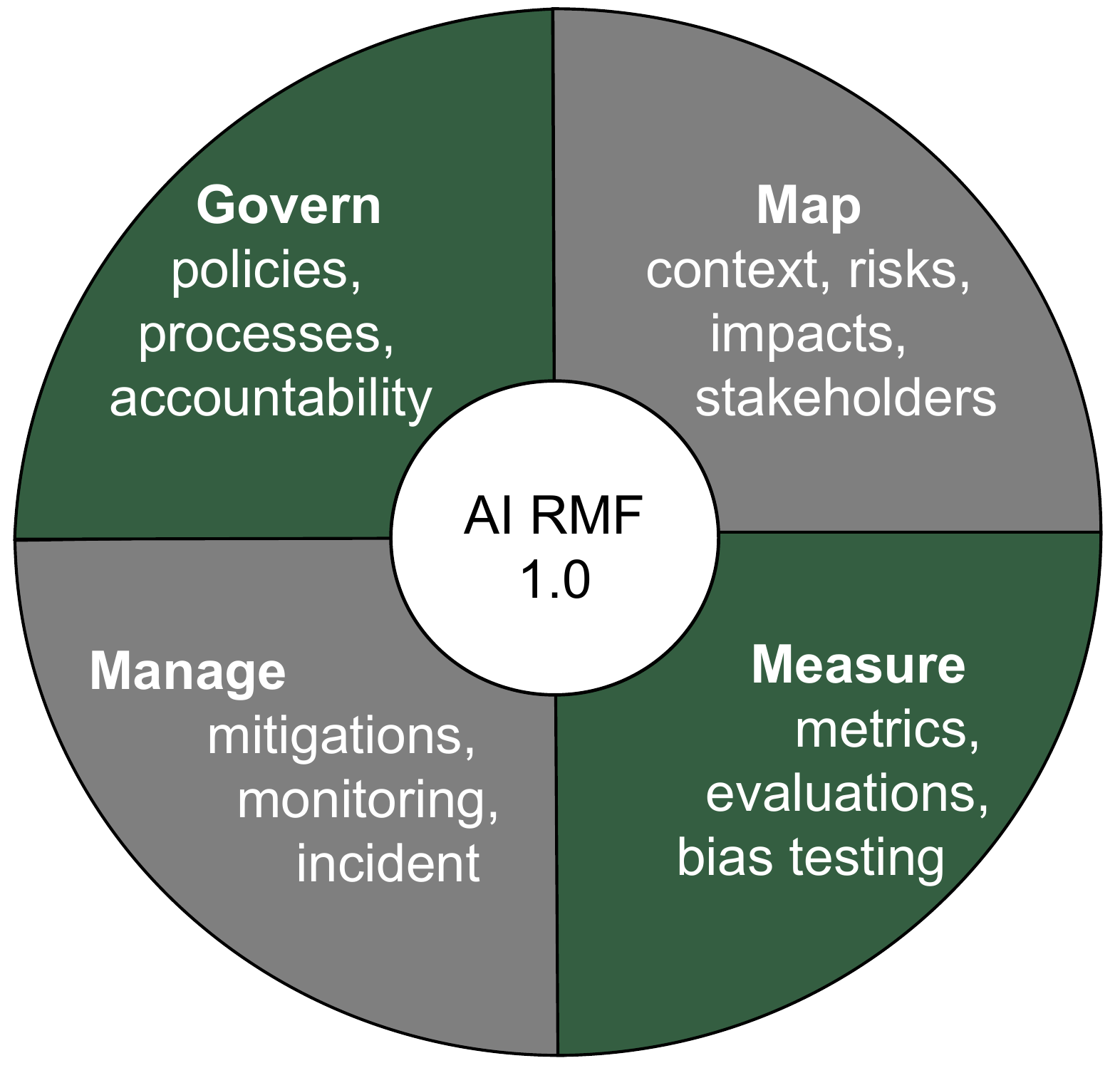}
    \caption{AI RMF}
    \label{fig:iec}
  \end{subfigure}
  \caption{Canonical illustrations of representative governance frameworks: (a) RAMI 4.0 model emphasizing layered architecture, (b) IDS connectors enabling sovereign data exchange, and (c) NIST AI RMF's four core functions. Note that none explicitly models trust propagation dynamics across communication networks.}
  \label{fig:gov-figures}
\end{figure*}

\begin{table*}[!t]
    \centering
    \caption{Comparison Between This Work and Existing Survey/Review Papers}
    \label{tab:compare_surveys}
    \resizebox{\textwidth}{!}{
    \begin{tabular}{lccccccc}
    \toprule
    \textbf{Survey / Review} & \textbf{Domain} & \textbf{Data Gov.} & \textbf{Service Gov.} & \textbf{Knowledge Gov.} & \textbf{Network/Comm. Awareness} & \textbf{Trust Dimensions} & \textbf{Cross-layer Propagation} \\
    \midrule
    Serrano et al.~\cite{s1} & Data governance & \checkmark & -- & -- & -- & Limited & -- \\
    Nan et al.~\cite{s2} & Microservice gov. & -- & \checkmark & -- & Partial (IT-centric) & -- & -- \\
    Xue et al.~\cite{s3} & Knowledge graphs & -- & -- & \checkmark & -- & -- & -- \\
    Gadekallu et al.~\cite{s4} & Trustworthy AI & Partial & -- & -- & -- & Strong (AI-centric) & -- \\
    De et al.~\cite{s5} & IIoT security & Partial & Partial & -- & Security-focused & Security-only & Partial \\
    Multi-party Comm.~\cite{10976591} & Secure computation & -- & -- & -- & Protocol-level & Privacy/Security & -- \\
    \midrule
    \textbf{This Work} & \textbf{Networked Industrial Gov.} & \textbf{\checkmark} & \textbf{\checkmark} & \textbf{\checkmark} & \textbf{\checkmark (Comm.-aware)} & \textbf{5D Unified} & \textbf{\checkmark (End-to-End)} \\
    \bottomrule
    \end{tabular}
    }
\end{table*}

In contemporary networked industrial systems, data, services, and knowledge jointly determine operational correctness, yet their governance remains critically fragmented. Data serve as the factual substrate~\cite{Wang2022DataGovernance}, services operationalize logic through automated workflows, and knowledge encodes domain semantics for reasoning~\cite{smcrina,meifang}. However, traditional data governance focuses on static quality and compliance, ignoring how network-induced delays or packet loss degrade downstream utility. Service governance enforces runtime SLAs but lacks semantic awareness of data provenance or model validity. Knowledge governance prioritizes interpretability yet remains disconnected from real-time operational telemetry and dynamic network states. This siloed approach creates a \textit{semantic-temporal mismatch}: discrepancies among data freshness, service responsiveness, and knowledge validity propagate across networked boundaries, leading to silent failures that neither RAMI 4.0 nor AI RMF can prevent (Fig.~\ref{fig:gov-figures}). The challenge is exacerbated in cross-organizational settings, where heterogeneous communication protocols, incompatible metadata schemas, and opaque AI logic hinder interoperable trust establishment.

To address these gaps, in this work, we introduce \textsc{TRISK} (\textit{\underline{TR}usted \underline{I}ndustrial data-\underline{S}ervice-\underline{K}nowledge governance}) as a communication-aware governance framework for designing and evaluating trustworthy networked industrial systems. Unlike descriptive reference architectures that map components without operational trust semantics, or isolated surveys that categorize techniques within single layers, \textsc{TRISK} prescribes explicit inter-layer contracts and feedback mechanisms that account for network dynamics. For instance, in predictive maintenance, a conventional stack might deploy accurate models fed by clean data via reliable services. Yet if the model was trained on a different machine variant and the network cannot deliver timely context metadata, a silent failure occurs. \textsc{TRISK} mandates that services validate incoming data metadata against model assumptions \textit{before} execution, that knowledge modules expose operational constraints, and that violations trigger network-aware feedback loops to quarantine data or adapt service routing. This transforms trust from a static certificate into a dynamic, runtime-enforced property resilient to communication uncertainties.

\begin{figure*}[!t] 
    \centering 
    \includegraphics[width=0.88\linewidth]{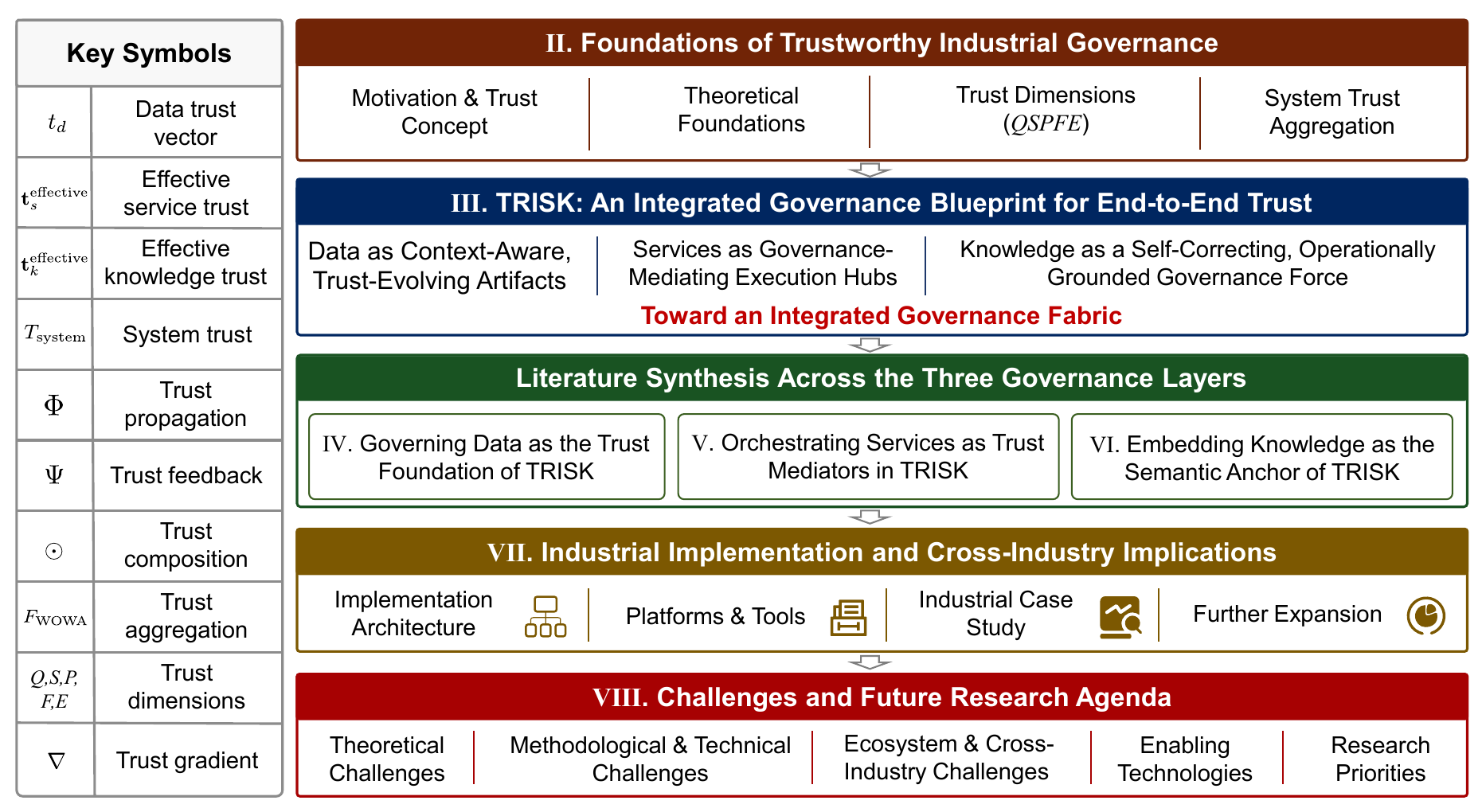} 
    \caption{Organization of the paper and logical progression of the \textsc{TRISK} framework. Section~\ref{found} establishes theoretical foundations including network-trust coupling. Section~\ref{sec:trisk} defines the \textsc{TRISK} blueprint. Sections~\ref{data}-\ref{knowledge} systematically review governance through the \textsc{TRISK} lens with emphasis on cross-layer propagation. Section~\ref{sec:implementation} evaluates industrial maturity, and Section~\ref{future} outlines a research agenda for integrated governance fabrics.} 
    \label{fig:orchestration} 
\end{figure*}

As summarized in Table~\ref{tab:compare_surveys}, this survey is the first to provide a unified, cross-layer examination of industrial governance through five interdependent trust dimensions, i.e., \textit{quality, security, privacy, fairness, and explainability}, while explicitly accounting for communication constraints. These dimensions reflect operational necessities: \textit{quality} ensures fidelity despite network impairments; \textit{security} mitigates cyber-physical threats across attack surfaces; \textit{privacy} respects data sovereignty in multi-tenant networks; \textit{fairness} prevents bias in resource-constrained edge environments; and \textit{explainability} enables causal attribution for safety-critical decisions. We construct the review corpus through a structured multi-source search covering IEEE Xplore, ACM Digital Library, ScienceDirect, SpringerLink, Scopus/Web of Science, Google Scholar, arXiv, and official standards or framework sources such as ISO/IEC, NIST, IEC, ISA, IDS, Gaia-X, and RAMI~4.0. Search terms cover industrial data governance, service governance, knowledge governance, IIoT trust, data spaces, edge-cloud orchestration, industrial AI governance, and knowledge-graph reasoning. We include peer-reviewed studies, standards, and technical reports with clear relevance to trustworthy industrial or cyber-physical governance, and code them by governance layer, trust dimension, communication awareness, validation type, maturity, and limitation. Our key contributions are as follows:
\begin{enumerate}
    \item A systematic taxonomy classifying industrial governance along three axes: i) governance objective (compliance- vs. quality- vs. service-aware), ii) architectural paradigm (centralized, federated, edge-embedded), and iii) enabling technology (knowledge graphs, zero-trust architectures, deterministic networking, and causal inference);
    \item A critical synthesis of 100+ representative works revealing how trust propagates or degrades across networked data-service-knowledge boundaries, identifying key deployability gaps including latency-sensitive policy enforcement, semantic-temporal mismatches, and lack of standardized cross-vendor governance interfaces;
    \item A forward-looking research agenda for Industry 5.0, outlining concrete open problems such as network-co-design trust mechanisms, lightweight edge governance, causal-aware data quality diagnosis under unreliable channels, and standardized APIs for interoperable trust fabrics.
\end{enumerate}

The remainder of this paper is structured as follows. Section~\ref{found} introduces foundational concepts of trustworthy governance in networked industrial systems. Section~\ref{sec:trisk} presents \textsc{TRISK} as a unified analytical lens, defining its core components and evaluation dimensions. Sections~\ref{data} through~\ref{knowledge} conduct a systematic review through the \textsc{TRISK} perspective: Section~\ref{data} examines data governance as the foundation of cross-layer trust; Section~\ref{service} investigates services as trust mediators orchestrating data and knowledge over dynamic networks; and Section~\ref{knowledge} explores knowledge systems as semantic anchors ensuring contextual coherence. Section~\ref{sec:implementation} synthesizes integration patterns and evaluates solution maturity. Building on this analysis, Section~\ref{future} articulates a research agenda centered on realizing an \textit{integrated governance fabric}, and Section~\ref{conclusion} concludes this work. Fig.~\ref{fig:orchestration} illustrates the logical flow.

\section{Foundations of Trustworthy Industrial Governance} \label{found}

Trustworthy governance in industrial intelligence cannot be reduced to a mere extension of enterprise data management or generic AI assurance. It must be rooted in the cyber-physical reality of networked production systems, where digital decisions actuate physical processes and trust failures propagate across organizational, temporal, and communication boundaries with tangible consequences. This section establishes the conceptual bedrock for a unified governance paradigm by: (i) characterizing the distinctive traits of industrial data and services that couple trust to network dynamics; (ii) formalizing a multidimensional trust model with layer-specific semantics grounded in information, decision, and game theory; (iii) presenting a unified, communication-aware trust aggregation framework enabling computable reasoning; and (iv) diagnosing why existing frameworks fail to bridge the network-governance semantic gap. These foundations motivate the integrative, network-co-design approach of \textsc{TRISK}, introduced in Section~\ref{sec:trisk}.

\subsection{Distinctive Characteristics of Networked Industrial Systems}
Industrial environments impose constraints that fundamentally differentiate their governance from conventional IT/cloud systems. These arise from the tight coupling between digital logic, physical operations, and deterministic communication networks. Specifically:

\begin{itemize}
    \item \emph{Multimodal Heterogeneity with Network-Induced Uncertainty}: Industrial systems ingest streams from sensors, PLCs, logs, and supply-chain records, each with distinct formats and sampling rates. Crucially, network jitter, packet loss, and protocol translation introduce additional uncertainty that directly degrades data fidelity and semantic alignment~\cite{11184582}.
    
    \item \emph{Network-Dependent Temporal Fidelity}: Control loops operate under strict timing constraints. Trust is not merely about data correctness but about \textit{timely correctness}; communication latency or scheduling delays can render otherwise valid data or decisions untrustworthy or even hazardous.
    
    \item \emph{Cross-Organizational Sovereignty over Shared Networks}: Value chains involve OEMs, suppliers, and integrators operating sovereign infrastructures over shared or federated networks. Governance must reconcile divergent compliance regimes while maintaining interoperable trust semantics across administrative and network domains.
    
    \item \emph{Coupled Physical Risk Amplified by Network Failures}: Misgovernance manifests as equipment damage or safety incidents. Network partitions, denial-of-service, or silent data corruption can trigger cascading physical failures that are indistinguishable from logical errors without network-aware provenance.
    
    \item \emph{Long-Lived Workflows over Evolving Network Topologies}: Industrial processes span decades, undergoing software updates, hardware retrofits, and network migrations (e.g., fieldbus to TSN). Governance must maintain consistent trust semantics across heterogeneous and evolving communication substrates.
\end{itemize}

These traits demand governance mechanisms that treat \textit{communication} not as a transparent pipe but as a first-class trust dimension. Trust in networked industrial systems is thus a dynamic, context-dependent condition requiring continuous verification across data, service, knowledge, \textit{and network} layers.

\subsection{Theoretical Foundations of Multi-Dimensional Trust}
\label{sec:theory}

Trust in industrial systems is inherently multi-dimensional and subject to fundamental trade-offs arising from cyber-physical-network coupling. Drawing from information theory, decision theory, and game theory, we establish underpinnings for the five trust dimensions (QSPFE) and their interdependencies, now explicitly incorporating communication constraints.

\paragraph{Information-Theoretic Trade-offs. Quality vs. Explainability under Latency Constraints} Quality ($Q$) minimizes divergence between observations and true states; explainability ($E$) maximizes mutual information between decisions and causal factors. Complex models achieve high $Q$ but low $E$. In networked settings, this trade-off is exacerbated by latency: high-fidelity explanations often require transmitting large metadata payloads, conflicting with real-time bandwidth limits. Thus, the $Q$-$E$ frontier is dynamically shaped by available communication capacity.

\paragraph{Security-Privacy Dilemma in Distributed Networked Ecosystems} Security ($S$) minimizes attack success probability; privacy ($P$) limits information disclosure. Aggressive security monitoring increases data collection (raising privacy risk), while strong privacy (e.g., differential privacy) obscures anomalies needed for intrusion detection. In cross-organizational networks, this dilemma intensifies: partners must share operational telemetry for collective security without exposing proprietary process details, requiring cryptographic and protocol-level co-design.

\paragraph{Fairness-Quality Tension under Resource and Bandwidth Scarcity} Fairness ($F$) ensures equitable resource allocation. Optimizing $F$ (e.g., equal maintenance scheduling) may divert resources from high-priority assets, reducing local $Q$. In network-constrained edge environments, this extends to bandwidth allocation: fair access may starve critical control flows, creating a tripartite $F$-$Q$-\textit{Bandwidth} trade-off requiring joint optimization.

\paragraph{Latency-Explainability Trade-off (New for Networked CPS)} Beyond classical $Q$-$E$, networked systems face a direct tension between explanation richness and timeliness. High-fidelity explanations (high $E$) often require aggregating distributed context, increasing latency and risking violation of real-time deadlines (low effective $Q$). This necessitates adaptive explanation strategies that scale fidelity to available network headroom.

These perspectives establish that QSPFE dimensions are coupled through mathematical trade-offs intrinsic to networked cyber-physical systems. \textsc{TRISK} must navigate these trade-offs in a principled, communication-aware manner.

\subsection{Five-Dimensional Trust Model: Layer-Specific Formalization with Network Coupling}
\label{sec:qspe_layers}

Building on Section~\ref{sec:theory}, we formalize how the five trust dimensions (QSPFE) manifest across data, service, and knowledge layers, with explicit coupling to network state. Each metric is defined as a computable function of observable attributes; forward references indicate operationalization in subsequent governance mechanisms (see Section \ref{data}).

\subsubsection{Data Layer. Trust as Measurable Attributes Modulated by Network State}
At the data layer, trust dimensions are quantified as attributes of industrial records degraded by network impairments:
\begin{itemize}
    \item \textit{Quality} ($Q_d$): Signal fidelity adjusted for network-induced uncertainty. Let $x(t)$ denote the sensor measurement at time $t$, $x_\text{true}(t)$ the corresponding ground-truth physical value, and $\sigma_n^2$ the variance of network-induced error (e.g., from packet loss or jitter). Then,
    \begin{equation}
        Q_d = 1 - \frac{\mathbb{E}[(x(t) - x_\text{true}(t))^2] + \sigma_n^2}{\mathbb{E}[x_\text{true}(t)^2]},
    \end{equation}
    where $\mathbb{E}[\cdot]$ denotes expectation over a sliding window.
    
    \item \textit{Security} ($S_d$): Integrity assurance against unauthorized access and network-layer tampering. Let $p_\text{breach}$ be the probability of logical breach and $p_\text{net\_tamper}$ the probability of network-level data manipulation. Assuming independence,
    \begin{equation}
        S_d = 1 - \big(p_\text{breach} + p_\text{net\_tamper} - p_\text{breach} \cdot p_\text{net\_tamper}\big),
    \end{equation}
    aligned with IEC 62443 security zones.
    
    \item \textit{Privacy} ($P_d$): Strength of differential privacy protection. For a mechanism satisfying $\epsilon$-differential privacy with privacy budget $\epsilon > 0$,
    \begin{equation}
        P_d = e^{-\epsilon},
    \end{equation}
    linked to IDS usage policies.
    
    \item \textit{Fairness} ($F_d$): Statistical parity across heterogeneous asset groups. Let $A \in \{0,1\}$ be a binary attribute (e.g., machine type) and $Y \in \{0,1\}$ the data collection outcome. Then,
    \begin{equation}
        F_d = 1 - \big| \Pr(Y=1 \mid A=0) - \Pr(Y=1 \mid A=1) \big|,
    \end{equation}
    extended to network-access fairness.
    
    \item \textit{Explainability} ($E_d$): Metadata completeness for traceability. Let $m$ be the total number of required metadata fields and $m_\text{filled}$ the number actually populated. Then,
    \begin{equation}
        E_d = \frac{m_\text{filled}}{m},
    \end{equation}
    enriched with network provenance tags.
\end{itemize}

\subsubsection{Service Layer: Trust as Runtime Guarantees over Dynamic Networks}
Service-layer trust translates into runtime guarantees modulated by network performance (see Section \ref{service}):
\begin{itemize}
    \item \textit{Quality} ($Q_s$): SLI adherence under latency constraints. Let $a_\text{obs}(\ell)$ be the observed accuracy under end-to-end latency $\ell$, $a_\text{target}$ the target accuracy, and $\ell_\text{max}$ the maximum tolerable latency. Define the indicator $\mathbb{I}[\cdot]$ as 1 if the condition holds and 0 otherwise. Then,
    \begin{equation}
        Q_s = \frac{a_\text{obs}(\ell)}{a_\text{target}} \cdot \mathbb{I}[\ell \leq \ell_\text{max}],
        \label{eq_001}
    \end{equation}
    forming the basis for adaptive SLA enforcement.
    
    \item \textit{Security} ($S_s$): Attack surface reduction. Let $\text{AS}_\text{initial}$ and $\text{AS}_\text{current}$ denote the initial and current attack surface areas (measured in exploitable entry points). Then,
    \begin{equation}
        S_s = 1 - \frac{\text{AS}_\text{current}}{\text{AS}_\text{initial}},
    \end{equation}
    integrated with zero-trust micro-segmentation.
    
    \item \textit{Privacy} ($P_s$): Data minimization adherence. Let $D_\text{accessed}$ be the set of data fields actually accessed by a service and $D_\text{required}$ the set strictly necessary for its function. Then,
    \begin{equation}
        P_s = 1 - \frac{|D_\text{accessed} \setminus D_\text{required}|}{|D_\text{accessed}|},
    \end{equation}
    enforced through policy-aware routing.
    
    \item \textit{Fairness} ($F_s$): Equitable response time allocation. For $n$ concurrent clients with response times $\{t_i\}_{i=1}^n$, 
    \begin{equation}
        F_s = 1 - \frac{\max_i(t_i) - \min_i(t_i)}{\max_i(t_i) + \min_i(t_i)},
    \end{equation}
    linked to TSN stream reservation.
    
    \item \textit{Explainability} ($E_s$): Audit trail completeness. Let $J$ be the set of justifications provided and $J_\text{required}$ the set mandated by governance policy. Then,
    \begin{equation}
        E_s = \frac{|J \cap J_\text{required}|}{|J_\text{required}|},
    \end{equation}
    augmented with network event logs.
\end{itemize}

\subsubsection{Knowledge Layer. Trust as Epistemic Confidence with Deployment Context}
Knowledge-layer trust reflects confidence in AI models conditioned on deployment context (see Section \ref{knowledge}):
\begin{itemize}
    \item \textit{Quality} ($Q_k$): Predictive accuracy combined with calibration. Let $\text{acc}$ be the model accuracy on a held-out test set and $\text{CE}$ the expected calibration error. Then,
    \begin{equation}
        Q_k = \text{acc} \cdot (1 - \text{CE}),
    \end{equation}
    validated against operational context.
    
    \item \textit{Security} ($S_k$): Adversarial robustness. Let $\text{acc}_\text{clean}$ and $\text{acc}_\text{adv}$ be accuracies on clean and adversarially perturbed inputs, respectively. Then,
    \begin{equation}
        S_k = \frac{\text{acc}_\text{adv}}{\text{acc}_\text{clean}},
    \end{equation}
    tested under network-injected perturbations.
    
    \item \textit{Privacy} ($P_k$): Resistance to membership inference attacks. Let $\text{MI}_\text{risk} \in [0,1]$ be the estimated success probability of such an attack. Then,
    \begin{equation}
        P_k = 1 - \text{MI}_\text{risk},
    \end{equation}
    assessed in federated learning setups.
    
    \item \textit{Fairness} ($F_k$): Equal opportunity across protected groups. Let $\text{TPR}_{G_0}$ and $\text{TPR}_{G_1}$ be true positive rates for groups $G_0$ and $G_1$. Then,
    \begin{equation}
        F_k = 1 - \big| \text{TPR}_{G_0} - \text{TPR}_{G_1} \big|,
    \end{equation}
    monitored for drift due to network-induced bias.
    
    \item \textit{Explainability} ($E_k$): Explanation fidelity. Let $f(x)$ be the model prediction and $g(x)$ the generated explanation for input $x$. Then,
    \begin{equation}
        E_k = 1 - \frac{\mathbb{E}_x\big[|f(x) - g(x)|\big]}{\max_x |f(x)|},
    \end{equation}
    adapted to operator bandwidth constraints.
\end{itemize}

\subsubsection{Cross-Layer Trust Propagation with Network State Integration} 
Trust propagates through inter-layer dependencies modulated by network state $\mathbf{n} = [n_\text{latency}, n_\text{jitter}, n_\text{loss}, n_\text{security}]^\top$, where each component is normalized to $[0,1]$. The \textit{network-aware propagation function} $\Phi$ defines effective service trust as
\begin{equation}
    \mathbf{t}_s^\text{effective} = \Phi(\mathbf{t}_s, \mathbf{t}_d, \mathbf{n}) = \mathbf{t}_s \odot \mathbf{t}_d \odot \boldsymbol{\gamma}(\mathbf{n}),
\end{equation}
where $\mathbf{t}_s, \mathbf{t}_d \in [0,1]^5$ are service and data trust vectors, $\odot$ denotes element-wise multiplication, and $\boldsymbol{\gamma}: [0,1]^4 \rightarrow [0,1]^5$ is a network degradation mapping (e.g., high $n_\text{latency}$ reduces $Q_s$ and $E_s$). Knowledge trust similarly inherits from training data $\mathbf{t}_d^\text{train}$, deployment services $\mathbf{t}_s^\text{deploy}$, and deployment network $\mathbf{n}^\text{deploy}$:
\begin{equation}
    \mathbf{t}_k^\text{effective} = \mathbf{t}_k \odot \mathbf{t}_d^\text{train} \odot \mathbf{t}_s^\text{deploy} \odot \boldsymbol{\gamma}(\mathbf{n}^\text{deploy}).
\end{equation}
Conversely, the \textit{feedback operator} $\Psi$ updates upstream trust based on downstream gradients and network changes:
\begin{equation}
    \mathbf{t}_\text{upstream}^{(t+1)} = \Psi \big(\mathbf{t}_\text{upstream}^{(t)}, \nabla T_\text{downstream}, \Delta \mathbf{n} \big),
\end{equation}
where $\nabla T_\text{downstream}$ is the trust gradient from the downstream layer and $\Delta \mathbf{n}$ captures network state changes. For instance, persistent latency spikes ($\Delta n_\text{latency} > 0$) may trigger $Q_d$ downgrades or re-routing, formalizing end-to-end trust as a network-contingent, continuously adapted property.

\subsection{A Unified, Communication-Aware Trust Aggregation Framework}
\label{sec:aggregation}

We present a unified framework for aggregating multi-dimensional, multi-layer trust into holistic scores, serving as the mathematical backbone of \textsc{TRISK} with embedded network awareness.

\subsubsection{Properties of Trust Aggregation Functions}
Any valid aggregation $F: [0,1]^n \rightarrow [0,1]$ must satisfy four axiomatic properties: \textit{boundedness} ($0 \leq F(\mathbf{t}) \leq 1$), \textit{monotonicity} ($\mathbf{t}_1 \leq \mathbf{t}_2 \Rightarrow F(\mathbf{t}_1) \leq F(\mathbf{t}_2)$), \textit{idempotence} ($F(c, \dots, c) = c$), and \textit{weak compensation} (high values partially compensate low ones, but $F(\mathbf{t}) < 1$ if any $t_i < 1$). These ensure intuitive, non-degenerate trust scoring.

\subsubsection{Weighted Ordered Weighted Averaging (WOWA) Operators}
We adopt WOWA operators to simultaneously capture dimension importance and compensation behavior. Given trust vector $\mathbf{t} = (t_1, \dots, t_n)$ and importance weights $\mathbf{w} = (w_1, \dots, w_n)$ with $\sum_{i=1}^n w_i = 1$, the WOWA operator is defined as
\begin{equation}
    F_\text{WOWA}(\mathbf{t}) = \sum_{i=1}^n q_i \cdot t_{\sigma(i)},
    \label{eq_wowa}
\end{equation}
where $\sigma$ is a permutation sorting $\mathbf{t}$ in non-increasing order, and coefficients $q_i$ derive from a linguistic quantifier $Q: [0,1] \rightarrow [0,1]$ via
\begin{equation}
    q_i = Q\left(\sum_{j=1}^i w_{\sigma(j)}\right) - Q\left(\sum_{j=1}^{i-1} w_{\sigma(j)}\right).
\end{equation}
In safety-critical modes, $Q(r) = r^\alpha$ ($\alpha > 1$) emphasizes weakest dimensions; in operational modes, $Q(r) = r$ recovers the weighted mean for balanced optimization.

\subsubsection{Multi-Layer Trust Composition}
Trust composition follows a hierarchical two-stage process:
\begin{enumerate}
    \item \textit{Intra-layer aggregation} computes layer-specific trust scores via $T^{(l)} = F_\text{WOWA}^{(l)}(\mathbf{t}^{(l)})$ for each layer $l \in \{d, s, k\}$.
    \item \textit{Inter-layer aggregation} synthesizes these into system trust via $T_\text{system} = F_\text{WOWA}^\text{layers}(T^{(d)}, T^{(s)}, T^{(k)})$, where inter-layer weights reflect operational criticality and network dependency.
\end{enumerate}

\subsubsection{Network-Aware Dynamic Weight Adaptation}
To embed communication awareness, weights adapt to network state $\mathbf{n}$ via
\begin{equation}
    \mathbf{w}^{(t+1)} = \mathbf{w}^{(t)} + \eta \cdot \nabla_{\mathbf{w}} U(\mathbf{t}, \mathbf{w}, \mathbf{n}),
\end{equation}
where $\mathbf{w}^{(t)}$ is the weight vector at time $t$, $\eta > 0$ is an adaptation rate, and $U(\mathbf{t}, \mathbf{w}, \mathbf{n})$ is a utility function balancing safety, efficiency, compliance, and network utilization. During congestion ($n_\text{loss} \uparrow$), $U$ may increase weight on $Q_d$/$S_d$ while decreasing $E_d$, enabling context-responsive trust aggregation.

\subsection{Systemic Limitations of Existing Governance Frameworks: The Network-Governance Gap}
Despite growing standards, no framework provides unified, network-aware trust propagation across data-service-knowledge layers. 

\begin{enumerate}
    \item \textit{Architectural Reference Models} (e.g., RAMI~4.0~\cite{RAMI}, IIRA~\cite{IIRA}, AAS~\cite{AAS}) offer structural blueprints and syntactic interoperability but are descriptive, not operational. They define components, not computable trust metrics or runtime enforcement logic across OT/IT/network boundaries.
    
    \item \textit{Data-Space Initiatives} (e.g., IDS~\cite{IDS}, Gaia-X~\cite{Gaia-X}) enable policy-aware cross-org data exchange but remain transactional and data-centric. They lack semantics to constrain downstream service consumption or knowledge inference, leaving orchestration ungoverned and network conditions unaccounted for in policy enforcement.
    
    \item \textit{Security and Privacy Frameworks} (e.g., NIST Privacy~\cite{NIST}, IEC/ISA~62443~\cite{IEC62443_3_2_2020}) codify threat modeling and zonal isolation but are horizontal and reactive. They offer limited mechanisms to translate network-detected breaches into adaptive service/knowledge responses within real-time control loops.
    
    \item \textit{AI Governance Standards} (e.g., ISO/IEC~42001~\cite{ISOIEC42001_2023}, NIST AI RMF~\cite{NIST_AIRMF_2023}) address model lifecycle and bias but are agnostic to industrial constraints like communication latency, fail-safe actuation, or OT security zones. They provide scant support for runtime trust enforcement or network-conditioned provenance.
\end{enumerate}

In conclusion, existing frameworks either manage networks without trust semantics or manage trust without network awareness. None establishes a unified representational language for trust attributes with formal, network-modulated semantics, nor embeds closed-loop adaptation based on computable, communication-aware trust metrics. This fragmentation reflects a foundational gap: the absence of an approach treating trust as a first-class, computable property intrinsically woven into every layer of the \textit{networked} industrial intelligence stack. The next section introduces \textsc{TRISK}, bridging this gap by (i) semantically formalizing QSPFE with network-coupled mathematical foundations, (ii) enabling coordinated enforcement across layers via communication-aware aggregation, and (iii) anchoring the system in an end-to-end, network-provenanced backbone supporting closed-loop adaptation and auditability.

\section{\textsc{TRISK}: An Integrated Governance Framework for End-to-End Trust}
\label{sec:trisk}
The \textsc{TRISK} framework operationalizes the theoretical foundations established in Section~\ref{found}. While Section~\ref{sec:aggregation} formalized trust as a computable property through WOWA operators and propagation functions, \textsc{TRISK} provides the runtime substrate that enforces these calculations. \textsc{TRISK} treats the mathematical constructs ($\mathbf{t}^{(l)}$, $\Phi$, $F_\text{WOWA}$, $\Psi$) as first-class architectural entities. This section details how the framework translates the five-dimensional trust model into cohesive governance capabilities across data, service, and knowledge layers, bridging abstract theory and industrial implementation.

\subsection{Data as Context-Aware, Trust-Evolving Artifacts}
\label{sec:data_layer}
Grounded in the vector $\mathbf{t}_d = (Q_d, S_d, P_d, F_d, E_d)$ defined in Section~\ref{sec:qspe_layers}, \textsc{TRISK} redefines data from static artifacts into dynamic, context-aware entities. Rather than merely storing values, data objects encapsulate their trust state and evolve via runtime feedback, ensuring that system inputs satisfy the preconditions of the theoretical model.

\begin{itemize}
    \item \textit{Embedded Trust Context.} Every data record is augmented with a cryptographic envelope containing its real-time trust vector $\mathbf{t}_d$. This envelope serves as a machine-readable passport that travels with the data. For instance, in predictive maintenance, a sensor reading lacking a valid $Q_d$ score is deemed invalid by downstream services according to the propagation rule $\Phi$. This capability concretizes the abstract quality dimension at the source, enabling verifiable data ingestion.
    
    \item \textit{Bidirectional Trust Feedback via $\Psi$.} Implementing the feedback operator $\Psi$, the data layer supports a write-back mechanism. When downstream systems detect deviations (e.g., model prediction errors attributable to sensor drift), they trigger updates to upstream trust scores, such as reducing $Q_d$. This closes the loop between data quality and model performance: data trust influences service outcomes, and outcome errors retroactively refine data trust assessments, realizing the continuous adaptation formalized in Section~\ref{sec:qspe_layers}.
    
    \item \textit{Dynamic Trust Scoring.} Trust scores are recalculated in real-time via streaming analytics rather than assigned statically during ingestion. Addressing the long-lived workflow challenge identified in Section~\ref{found}, $\mathbf{t}_d$ values are continuously updated to reflect evolving operational contexts (e.g., machinery wear or network degradation). This ensures the data layer remains resilient to concept drift and maintains alignment with the current physical state.
\end{itemize}

These capabilities transform data from inert records into active participants in a distributed trust network, responsive to both interpretation and corrective action.

\begin{figure*}[!t] 
    \centering 
    \includegraphics[width=0.85\linewidth]{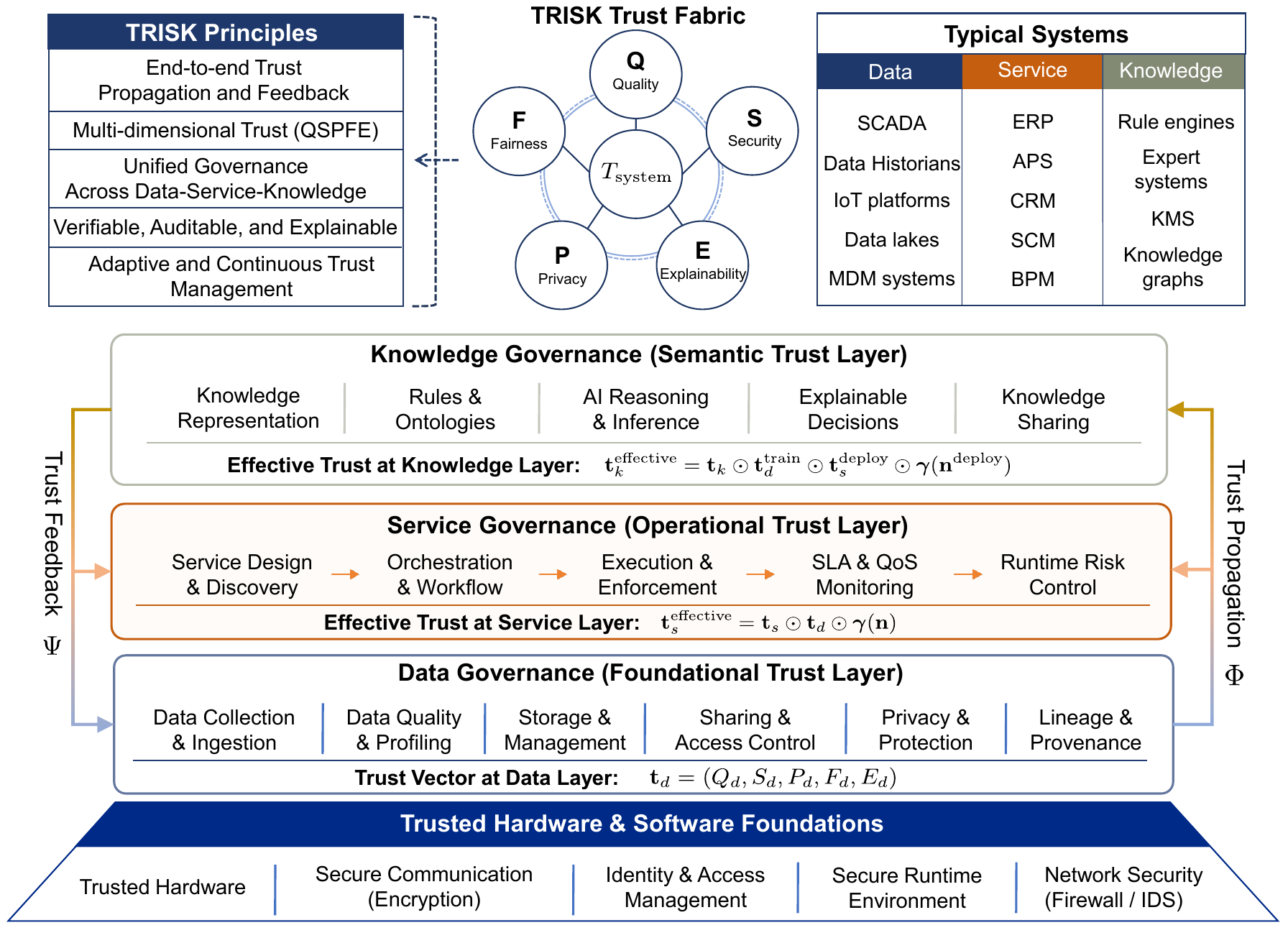} 
    \caption{\textsc{TRISK} Conceptual Architecture: Data, Service, and Knowledge governance planes are interwoven through the propagation function $\Phi$ and feedback operator $\Psi$, enabling end-to-end trust propagation and closed-loop adaptation. The Trusted Execution Foundation provides the hardware root of trust for all governance computations.} 
    \label{fig:Conceptual} 
\end{figure*}

\subsection{Services as Governance-Mediating Execution Hubs}
\label{sec:service_layer}
Building upon the service trust vector $\mathbf{t}_s$ and cross-layer propagation rules, services in \textsc{TRISK} function as governance mediators that enforce the mathematical constraints binding data to knowledge. In latency-sensitive industrial environments, this layer ensures that only trustworthy data flows are processed under dynamically adapted contracts.

\begin{itemize}
    \item \textit{Semantic Fitness Validation.} Before execution, services validate that incoming $\mathbf{t}_d$ satisfies the preconditions specified by the service's own trust contract $\mathbf{t}_s$. This architecturally enforces the element-wise multiplication ($\odot$) in $\Phi$. For example, a service requiring high privacy ($P_s$) for proprietary formulas rejects streams lacking requisite $P_d$ metadata, even if syntactically valid. This prevents the ``Garbage In, Gospel Out'' problem by making trust verification a prerequisite for computation.
    
    \item \textit{Governance Telemetry Exposure.} Services expose structured runtime telemetry (latency $\ell$, resource usage, SLA compliance) that directly parameterizes the aggregation function $F_\text{WOWA}$. This telemetry is not merely observational; it constitutes critical input for computing layer-specific trust $T^{(s)}$ and system-wide trust $T_\text{system}$. By logging assumption violations (e.g., $\ell > \ell_\text{max}$ degrading $Q_s$), services provide the evidentiary basis for knowledge-layer diagnostics.
    
    \item \textit{Adaptive Contract Execution.} Service contracts adapt based on aggregated trust scores, implementing the security-privacy trade-off resolution discussed in Section~\ref{sec:theory}. If $T_\text{system}$ drops below a safety threshold due to degraded security ($S_d$ or $S_s$), services automatically degrade functionality (e.g., switching from autonomous control to manual override). This enforces safety-first governance through mathematically grounded actuation constraints.
\end{itemize}

These capabilities render each service invocation an act of cross-layer alignment, elevating orchestration from a technical procedure to a core governance function.

\subsection{Knowledge as a Self-Correcting, Operationally Grounded Governance Force}
\label{sec:knowledge_layer}
The knowledge layer operationalizes epistemic confidence $\mathbf{t}_k$ and serves as the central intelligence for the governance fabric. Beyond model repository functions, it performs high-level reasoning and closed-loop adaptation across the entire system.

\begin{itemize}
    \item \textit{Multi-Layer Discrepancy Diagnosis.} Leveraging $T_\text{system}$ and component vectors $\mathbf{t}^{(l)}$, the knowledge system diagnoses root causes rather than symptoms. Unlike traditional AIOps, \textsc{TRISK} pinpoints whether failures stem from data noise (low $Q_d$), service latency (low $Q_s$), or model bias (low $F_k$). This unified diagnostic view addresses the systemic limitations of siloed monitoring by exploiting cross-layer dependencies formalized in $\Phi$.
    
    \item \textit{Bidirectional Knowledge-Data Binding via $\Psi$.} Enforcing the feedback loop $\Psi$, the knowledge layer retroactively adjusts training data trust $\mathbf{t}_d^\text{train}$ and model robustness $S_k$ upon detecting systemic errors. For instance, a predictive maintenance failure triggers not only model retraining but also trust label revision on historical data. This ensures the mathematical model evolves with operational experience and prevents error recurrence through provenance-aware learning.
    
    \item \textit{Self-Auditing Knowledge Artifacts.} All knowledge-layer decisions are logged with complete trust calculation provenance, including WOWA weights $\mathbf{w}$ and quantifier parameters $Q$. This creates an immutable audit trail explaining both the decision (e.g., line shutdown) and its mathematical justification (e.g., $S_d < 0.5$ triggering conjunctive aggregation with $\alpha > 1$). Such traceability is essential for regulatory compliance in safety-critical industries.
\end{itemize}

These capabilities elevate knowledge from a static decision engine to a dynamic governance agent that learns from and adapts to the operational ecosystem.

\subsection{Toward an Integrated Governance Fabric} 
\label{sec:integrated_fabric} 

The three capabilities above converge within a unified conceptual architecture that transforms fragmented governance into an integrated, multi-layer trust fabric. As depicted in Fig.~\ref{fig:Conceptual}, this fabric rests on five interlocking components that work in concert to enforce the \textsc{TRISK} paradigm.

At its base, a \textit{Trusted Execution Foundation} built on secure enclaves and hardware attestation ensures the integrity of all computations, providing the \textit{Root of Trust} necessary for the mathematical model to hold. Above this:
\begin{itemize} 
    \item[a)] The \textit{Data Governance Plane} manages the lifecycle of $\mathbf{t}_d$, ensuring data provenance and context-awareness. It acts as the memory of the system.
    \item[b)] The \textit{Service Governance Plane} acts as the operational nexus, enforcing the propagation function $\Phi$ and calculating $\mathbf{t}_s$. It is the muscle that executes workflows under governance constraints.
    \item[c)] The \textit{Knowledge Governance Plane} performs the final aggregation $F_\text{WOWA}$ to compute $T_\text{system}$, providing the holistic view. It is the brain that diagnoses issues and drives adaptation.
\end{itemize} 

Critically, the architecture visualizes the equations from Section~\ref{found} as physical data flows: validated data ($\mathbf{t}_d$) inform reliable service execution ($\mathbf{t}_s$); service outcomes shape knowledge confidence ($\mathbf{t}_k$); and knowledge contradictions feed back to refine data trust and service policies via $\Psi$. This closed-loop, bidirectional interweaving, where the output of one layer is the input for the trust calculation of another, is what distinguishes \textsc{TRISK} from conventional layered models. The result is a system capable of explaining not just \textit{what} they decide, but \textit{why it is trustworthy} based on the rigorous mathematical foundation laid out in the previous section. The following three sections conduct a literature synthesis through the \textsc{TRISK} lens, reviewing the key technologies underpinning data, service, and knowledge governance, and comparatively analyzing their objectives, mechanisms, strengths, and limitations in supporting trustworthy industrial intelligence.

\section{Governing Data as the Trust Foundation of TRISK} \label{data}

Data governance serves as the foundational anchor for constructing the initial data trust vector $\mathbf{t}_d = (Q_d, S_d, P_d, F_d, E_d)$ defined in Section~\ref{sec:qspe_layers}. In the \textsc{TRISK} framework, data are not static artifacts but dynamic carriers of trust evidence that propagate upward via $\Phi$ and receive feedback via $\Psi$. Therefore, trustworthy data governance must transcend traditional siloed management to become a cross-layer enabler. This section reviews representative studies through the \textsc{TRISK} lens, organizing the analysis around three core tensions that determine whether data mechanisms can effectively initialize and sustain end-to-end trust propagation.

\subsection{Industrial Data Characteristics as Drivers of Cross-Layer Trust Requirements}

The unique nature of industrial data imposes fundamental constraints on how $\mathbf{t}_d$ can be computed and propagated. We identify three characteristic-driven challenges that existing governance frameworks often fail to address holistically.

\subsubsection{Heterogeneity-Induced Semantic-Temporal Mismatch}
Industrial data span structured logs, semi-structured OPC-UA messages, and unstructured vibration signals~\cite{data}. This heterogeneity creates a \textit{semantic-temporal mismatch}: semantic alignment (e.g., unit normalization) often occurs at batch timescales, while trust propagation via $\Phi$ requires real-time consistency. For instance, a temperature reading mislabeled as ``$^{\circ}$F'' instead of ``$^{\circ}$C'' may pass syntactic validation but catastrophically degrade downstream service quality $Q_s$ and knowledge fairness $F_k$. Existing governance tools typically resolve semantics offline, leaving a temporal gap during which $\mathbf{t}_d$ is semantically invalid yet technically consumable, violating the monotonicity property of trust aggregation (Section~\ref{sec:aggregation}).

\subsubsection{Real-Time Dynamics vs. Static Trust Credentials}
Unlike batch analytics, industrial streams operate under strict latency constraints~\cite{critical}. However, most data governance mechanisms produce \textit{static trust credentials} (e.g., one-time quality certificates) that fail to capture the time-varying nature of $Q_d$ and $S_d$. When network jitter increases $\sigma_n^2$ in Eq.~(1), $Q_d$ must degrade instantaneously; static credentials cannot reflect this, causing $\Phi$ to propagate stale trust values. This discrepancy renders services unable to adapt their behavior (e.g., switching to fallback models) based on current data fidelity, breaking the closed-loop assumption of \textsc{TRISK}.

\subsubsection{Cross-Boundary Flows and Trust Discontinuity}
Modern operations span LIMS, MES, and cloud-edge infrastructures~\cite{interact,IOTziqi}, creating trust boundaries where schemas, policies, and update frequencies diverge. At these interfaces, $\mathbf{t}_d$ often undergoes \textit{trust discontinuity}: provenance metadata is stripped, privacy budgets $\epsilon$ are reset without composition accounting, or security contexts are re-established without inheriting upstream $S_d$. Without explicit propagation-aware synchronization, the element-wise multiplication in $\Phi$ receives incomplete or inconsistent $\mathbf{t}_d$ vectors, making cross-layer trust computation mathematically unsound regardless of individual system correctness.

\subsection{Core Mechanisms: Operationalizing $\mathbf{t}_d$ for Cross-Layer Propagation}

We analyze four mechanism families through their capability to compute, preserve, and propagate $\mathbf{t}_d$ components. Critically, we assess each against the \textsc{TRISK} requirement of supporting bidirectional trust flow.

\subsubsection{Quality Assurance and Standardization: Computing $Q_d$ and $E_d$ for Downstream Consumption}
Quality mechanisms directly instantiate $Q_d$ and $E_d$. Outlier detection and drift monitoring~\cite{10854807} provide the statistical basis for $Q_d$, while metadata schemas populate $E_d$. Digital twins enhance explainability by binding raw signals to domain semantics, enriching $E_d$ with physical context essential for knowledge-layer reasoning. AI-driven pipelines automate validation: blockchain-integrated agents verify completeness before anchoring~\cite{10402553}, and federated aggregation filters unreliable participants~\cite{11145130,10949664}.

Standardization frameworks (AutoDP, AutoFE~\cite{10909604}) reduce semantic ambiguity, indirectly supporting $F_d$ by ensuring consistent feature definitions across asset groups. Enterprise schema synchronization~\cite{11082505} maintains long-term coherence.

\textit{TRISK Gap Analysis:} Most quality mechanisms compute $Q_d$ in isolation without exposing it as a machine-readable attribute consumable by $\Phi$. Drift detectors signal degradation locally but rarely trigger upstream $\Psi$ feedback. Standardization efforts focus on syntactic interoperability rather than semantic trust preservation, leaving $E_d$ incomplete at cross-system boundaries. Consequently, services receive data with implicit rather than explicit quality guarantees, forcing them to assume worst-case $Q_d$ and underutilizing available trust evidence.

\subsubsection{Security and Privacy: Embedding $S_d$ and $P_d$ as Propagatable Attributes}
Security and privacy mechanisms instantiate $S_d$ and $P_d$. Preventive controls (AES, RBAC/ABAC~\cite{8972565,10854807,10539002}) bound $p_\text{breach}$, while zero-knowledge proofs enable verification without disclosure. Federated learning with differential privacy~\cite{11145130} provides formal $P_d$ guarantees, and blockchain auditing~\cite{10471193} enhances $S_d$ via tamper-evidence. Smart contracts (Cydon~\cite{8972565}) link access to verifiable obligations, transforming security from perimeter defense into an inspectable attribute. Threat modeling (STRIDE, LINDDUN~\cite{9973285}) and Web3 identity~\cite{10505933} embed compliance proactively.

\textit{TRISK Gap Analysis:} Current mechanisms treat $S_d$ and $P_d$ as binary gates (allow/deny) rather than continuous variables suitable for WOWA aggregation. Differential privacy budgets $\epsilon$ are rarely tracked across composition chains, making $P_d$ non-composable in multi-hop propagation. Security attestations lack temporal validity bounds, failing to capture dynamic $p_\text{net\_tamper}$. Thus, services cannot perform fine-grained trade-offs (e.g., accepting lower $P_d$ for higher $Q_d$ under emergency conditions), and knowledge systems cannot reason about cumulative privacy loss.

\subsubsection{Provenance and Accountability: Enabling $\Psi$ Feedback via Traceable $E_d$}
Provenance mechanisms provide the audit trail necessary for $\Psi$ to function. Metadata-driven lineage~\cite{10854807} populates $E_d$ and enables root-cause diagnosis. Blockchain ledgers~\cite{9712447,10596314} offer tamper-proof verification, while watermarking~\cite{11184582} and federated reputation~\cite{10949664} enable active accountability. RFID/IoT bridges~\cite{9841042} ground digital provenance in physical reality, and drift detectors (ADWIN, EDDM~\cite{10909604}) link data evolution to service performance changes.

\textit{TRISK Gap Analysis:} Provenance systems excel at retrospective auditing but lack real-time write-back capability required by $\Psi$. Lineage graphs record transformations but do not annotate trust deltas ($\Delta \mathbf{t}_d$) needed for gradient-based feedback. Reputation scores are typically scalar and unidimensional, incompatible with the five-dimensional $\mathbf{t}_d$ vector. Consequently, when knowledge layers detect errors, they cannot precisely attribute blame to specific $Q_d$, $S_d$, or $F_d$ degradations, nor can they retroactively adjust the correct component of upstream $\mathbf{t}_d$.

\subsubsection{Data Sharing and Marketplaces: Synthesizing $\mathbf{t}_d$ for Collaborative Intelligence}
Data marketplaces integrate QSPFE dimensions into collaboration frameworks. Policy-driven sharing via smart contracts~\cite{10402553} and consent management~\cite{10596314} automate compliance. Economic incentives (IV-TAOs~\cite{10415232}, reputation~\cite{9712447}) align contributor behavior with trustworthiness. Federated architectures~\cite{10949664} enable knowledge co-creation without raw data exposure, while IoT-cloud integration~\cite{9841042} supports time-sensitive exchange.

\textit{TRISK Gap Analysis:} Marketplace trust models are predominantly transactional rather than propagational. Trust scores serve pricing/ranking functions but are not designed as inputs to downstream $\Phi$. Multi-party computation ensures confidentiality but obscures the per-dimension trust decomposition needed for diagnostic reasoning. Incentive mechanisms optimize for local utility rather than global $T_\text{system}$, potentially encouraging high-$Q_d$ but low-$F_d$ contributions that degrade overall fairness. Thus, marketplaces generate isolated trust islands rather than a unified, propagatable trust fabric.

Table~\ref{tab:data_trisk_gap} synthesizes the cross-layer propagation capabilities and TRISK alignment gaps across all four mechanism families. The table reveals a systemic pattern: existing data governance mechanisms excel at local, static trust assurance but consistently lack the architectural interfaces required for real-time $\Phi$ consumption and $\Psi$ feedback. This gap analysis directly motivates the design principles articulated in Section~\ref{sec:integrated_fabric}.

\begin{table*}[htbp]
\centering
\renewcommand{\arraystretch}{0.8}
\caption{Cross-Layer Trust Propagation Capability of Data Governance Mechanisms}
\label{tab:data_trisk_gap}
\small
\begin{tabular}{p{3.4cm} p{3.8cm} p{3.4cm} p{5.5cm}}
\toprule
\textbf{Mechanism Family} & \textbf{$\mathbf{t}_d$ Components Supported} & \textbf{Propagation Strength} & \textbf{TRISK Alignment Gap} \\
\midrule
Quality \& Standardization & $Q_d$, $E_d$ & Computes metrics locally; limited real-time exposure to $\Phi$ & Metrics not packaged as machine-readable attributes; drift signals lack $\Psi$ write-back; semantic alignment lags real-time trust updates \\
\addlinespace
Security \& Privacy & $S_d$, $P_d$ & Binary enforcement dominant; continuous scoring rare & $S_d/P_d$ treated as gates not WOWA-compatible variables; $\epsilon$ non-composable across hops; attestation lacks temporal validity for dynamic $p_\text{net\_tamper}$ \\
\addlinespace
Provenance \& Accountability & $E_d$ (primary), all (diagnostic) & Retrospective audit strong; real-time feedback weak & Lineage lacks trust delta annotations; reputation is scalar/unidimensional; no precise attribution for $\Psi$ to adjust specific $\mathbf{t}_d$ components \\
\addlinespace
Sharing \& Marketplaces & All (synthesized) & Transactional trust scoring; propagational design absent & Scores optimized for pricing not $\Phi$ input; MPC obscures per-dimension decomposition; incentives misaligned with global $T_\text{system}$ optimization \\
\bottomrule
\end{tabular}
\end{table*}

\subsection{Toward Integrated Trust: Data Governance as the \textsc{TRISK} Foundation}

In \textsc{TRISK}, data governance is not a siloed function but the initial trust anchor from which cross-layer propagation begins. By operationalizing the five trust dimensions during data ingestion, transformation, and sharing, it creates a verifiable baseline upon which service execution and knowledge reasoning depend. High-quality, secure, and well-documented data enable services to operate reliably and allow knowledge systems to construct accurate, interpretable, and traceable representations. Conversely, the closed-loop design of \textsc{TRISK} ensures that trust violations detected at the service or knowledge layers, such as anomalous workflow outcomes or logically inconsistent inferences, can trigger data revalidation, lineage correction, or source quarantine. This bidirectional feedback prevents silent error accumulation and maintains systemic coherence. The reviewed data governance mechanisms constitute the upstream trust construction layer of \textsc{TRISK}: they determine the quality of $\mathbf{t}_d$, condition the reliability of subsequent trust propagation, and provide the evidential basis for feedback-driven correction.

Governing data as the foundation of cross-layer trust therefore means designing mechanisms that are not only technically robust but also semantically aware, temporally responsive, and architecturally coupled to downstream layers. Only through such intentional integration can industrial systems achieve the end-to-end trustworthiness demanded by Industry 5.0.

\section{Orchestrating Services as Trust Mediators in TRISK} \label{service}

Service governance constitutes the operational mediation layer that transforms data trust $\mathbf{t}_d$ into executable service trust $\mathbf{t}_s$, enforces knowledge-derived constraints, and generates runtime evidence for feedback adaptation via $\Psi$. In \textsc{TRISK}, services are not passive compute pipelines but active trust mediators that implement the propagation function $\Phi$ at runtime. This section reviews industrial service governance literature through the \textsc{TRISK} lens, examining how service characteristics shape cross-layer trust mediation and synthesizing four mechanism families by their capability to propagate, verify, and enforce trust across the data-service-knowledge continuum.

\subsection{Industrial Services as Cross-Layer Trust Mediators}

The architectural and operational demands of industrial services determine their capacity to bridge trust states between layers. We identify two fundamental tensions that existing orchestration frameworks often fail to resolve.

\subsubsection{Modular Composition vs. Fragmented Trust Contexts}
Service-oriented architectures encapsulate production, logistics, and maintenance functions as composable units~\cite{SOA}, enabling dynamic integration across cloud-edge and organizational boundaries. However, this modularity fragments trust: a workflow may combine services with heterogeneous provenance, reliability profiles, and policy constraints. The element-wise multiplication in $\Phi$ requires consistent $\mathbf{t}_d$ and $\mathbf{t}_s$ vectors, yet current orchestration engines validate syntactic compatibility without verifying semantic trust alignment. A service may accept syntactically valid data with degraded $Q_d$ or mismatched $P_d$, propagating trust violations downstream despite local correctness. Service governance must therefore mediate not only interface contracts but also trust vector coherence.

\subsubsection{Real-Time Execution vs. Static Trust Contracts}
Industrial services operate under stringent latency and safety constraints where failures directly impact physical processes~\cite{ser}. SLAs formalize trust expectations as quantifiable metrics~\cite{sla}, yet these contracts are typically static and scalar. When network state $\mathbf{n}$ changes (e.g., latency spike increasing $\ell$ in \eqref{eq_001}), $Q_s$ must degrade instantaneously; static SLAs cannot capture this dynamics, causing $\Phi$ to propagate stale trust values. Moreover, SLAs rarely decompose into the five-dimensional $\mathbf{t}_s$ vector required for WOWA aggregation. Services thus lack the granular trust signals needed for adaptive execution (e.g., switching fallback paths based on specific $S_s$ degradation rather than aggregate SLA breach), breaking the closed-loop assumption of \textsc{TRISK}.

\subsection{Core Mechanisms for Cross-Layer Trust Mediation}

We analyze four mechanism families by their capability to compute $\mathbf{t}_s$, enforce $\Phi$, and generate evidence for $\Psi$. Critically, we assess each against the \textsc{TRISK} requirement of bidirectional trust propagation.

\subsubsection{Service Lifecycle Management as Trust-Aware Orchestration}
Lifecycle governance embeds cross-layer awareness from deployment to decommissioning. Containerized microservices enable dynamic edge deployment~\cite{9918040}, while routing algorithms (CDRA, MA-OP) balance energy and continuity~\cite{10621029,hyuan}. Reinforcement learning adjusts scheduling based on real-time workload and trust metrics~\cite{10857627}, and telemetry-driven systems maintain SLOs through autonomous configuration~\cite{9612603}. Intent-based frameworks translate operational goals into executable configurations using ontologies~\cite{11068168,11036691}, preserving semantic alignment.

\textit{TRISK Gap Analysis:} Lifecycle mechanisms optimize resource efficiency and availability but rarely expose trust state as a first-class orchestration variable. RL reward functions typically maximize throughput or latency, not $\mathbf{t}_s$ components. Intent-to-configuration mappings lack trust vector decomposition, making it impossible to verify whether deployed services satisfy specific $F_s$ or $E_s$ requirements. Consequently, lifecycle management produces operationally efficient but trust-opaque services, forcing downstream layers to infer $\mathbf{t}_s$ indirectly rather than consuming it as explicit metadata.

\subsubsection{Reliability Assurance and SLA Enforcement as Contractual Trust Anchors}
Reliability mechanisms anchor trust in verifiable performance guarantees. Dynamic redundancy and resilient chaining ensure continuity under partial failures~\cite{9399165,9941499}, while GNN-based prediction enables anticipatory reallocation~\cite{11138027}. Hierarchical SLA frameworks (NFV-MANO, 5GT-MON) coordinate compliance across layers~\cite{9459426,9508404}, and smart contracts encode SLA terms into executable logic~\cite{10102926,9841004}. Some contracts reference data provenance or knowledge rules, linking SLA enforcement to cross-layer constraints.

\textit{TRISK Gap Analysis:} SLAs remain predominantly scalar and unidimensional, incompatible with the five-dimensional $\mathbf{t}_s$ vector required for WOWA aggregation. Smart contracts evaluate binary compliance (pass/fail) rather than continuous trust degrees, preventing fine-grained trade-offs (e.g., accepting lower $Q_s$ for higher $S_s$ under emergency). Reliability predictions forecast failure probability but do not decompose into specific $\mathbf{t}_s$ component degradations. Thus, services cannot serve as precise trust mediators; they either fully execute or fully fail, lacking the graduated adaptation needed for resilient $\Phi$ propagation.

\subsubsection{Policy-Driven Governance as Semantic and Ethical Mediation}
\begin{figure}[htbp]
    \centering
    \includegraphics[width=0.9\linewidth]{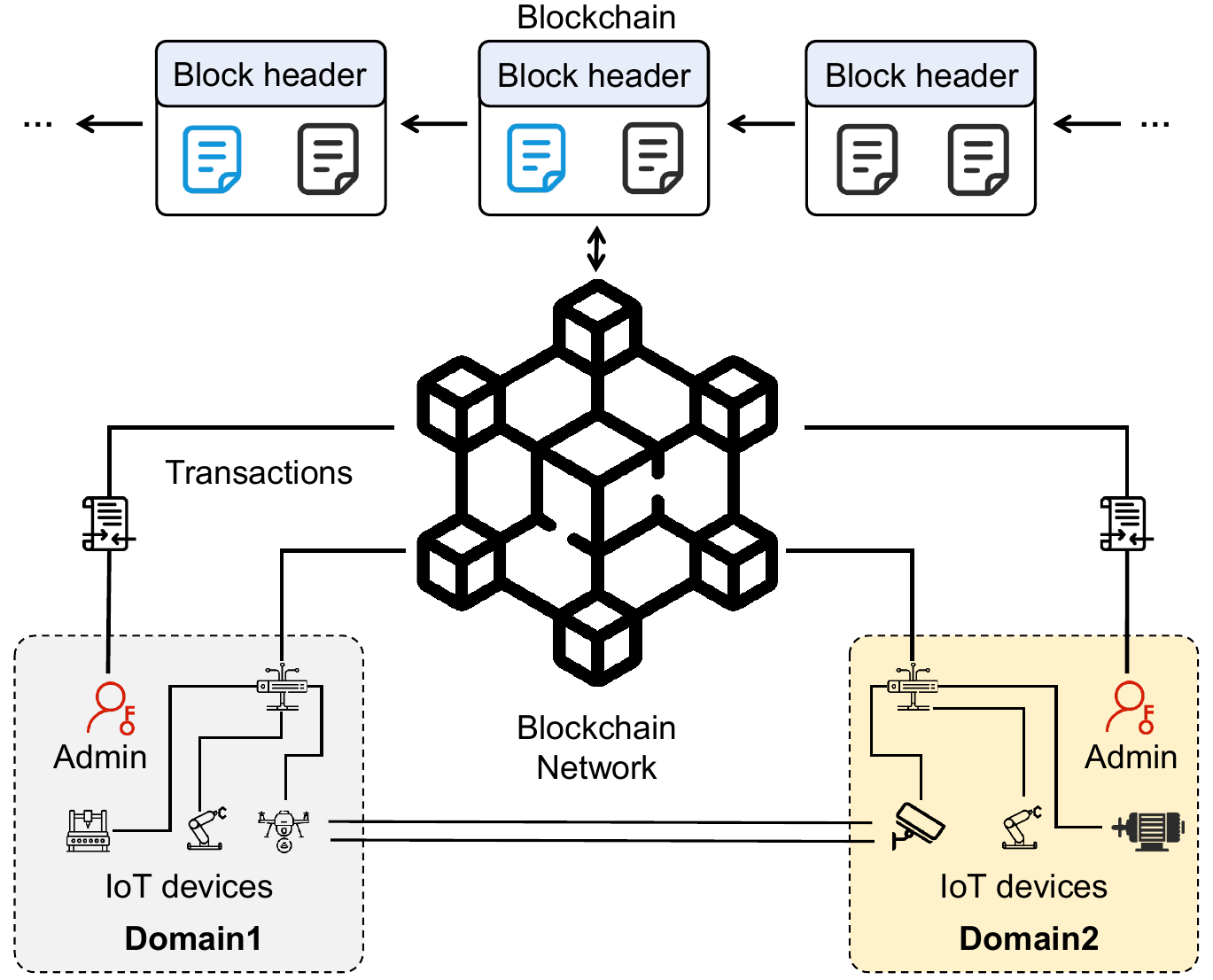}
    \caption{Blockchain-based access policy enforcement architecture~\cite{9786793}, integrating IoT terminals, blockchain networks, and access domains for auditable policy mediation.}
    \label{fig:blockchain_model}
\end{figure}

Policy models regulate service behavior under complex constraints. Rule-based systems (IEC~61499) maintain coherence across distributed microservices~\cite{9918040}, while AHP-based allocators balance multi-criteria objectives~\cite{9419855}. Blockchain-based access control embeds conditional policies into tamper-proof transactions~\cite{9786793} (Fig.~\ref{fig:blockchain_model}). Ontology-driven frameworks map human intents to machine-executable constraints~\cite{11036691}, and formal methods (CTL, $\pi$-calculus) verify policy safety~\cite{9864250,10660508}.

\textit{TRISK Gap Analysis:} Policy mechanisms excel at deterministic constraint enforcement but lack dynamic trust propagation capability. Rules are typically boolean (allow/deny) rather than continuous, incompatible with WOWA's weak compensation property. Blockchain policies ensure immutability but do not track trust deltas ($\Delta \mathbf{t}_s$) needed for $\Psi$ feedback. Formal verification proves safety properties statically but cannot validate runtime trust evolution under changing $\mathbf{n}$. Consequently, policies act as rigid gates rather than adaptive trust mediators, unable to support the graduated, context-responsive enforcement required by $\Phi$.

\subsubsection{Runtime Verification as Dynamic Trust Feedback Generation}
Runtime mechanisms ensure compliance during execution and generate evidence for cross-layer feedback. Temporal logic monitors evaluate event streams against safety properties~\cite{9411895}, while FPGA-accelerated verification reduces latency for control-critical loops~\cite{10726601}. ML models predict violations preemptively~\cite{10745554}. Decentralized auditing via blockchain logs ensures tamper-proof accountability~\cite{9841004,10419037}, and security-aware frameworks link logs to anomaly detection~\cite{9775180}. Runtime violations trigger data revalidation or rule refinement, closing the \textsc{TRISK} loop.

\textit{TRISK Gap Analysis:} Runtime verification generates abundant telemetry but rarely structures it as $\mathbf{t}_s$ components consumable by $\Psi$. Monitors detect violations but do not attribute them to specific $Q_s$, $S_s$, or $F_s$ degradations. Audit logs record events retrospectively but lack real-time write-back interfaces for upstream trust updates. Predictive models forecast breaches but do not output trust vectors suitable for WOWA aggregation. Thus, services produce rich but unstructured evidence, forcing knowledge layers to perform costly inference rather than consuming pre-computed $\mathbf{t}_s$ feedback.

\begin{table*}[t]
\centering
\footnotesize
\renewcommand{\arraystretch}{1.1}
\caption{Cross-Layer Trust Mediation Capability of Service Governance Mechanisms}
\label{tab:service_trisk_gap}
\begin{tabular}{@{}p{2.6cm} p{3.4cm} p{3.8cm} p{6.5cm}@{}}
\toprule
\textbf{Mechanism Family} & \textbf{$\mathbf{t}_s$ Components Supported} & \textbf{Propagation Strength} & \textbf{TRISK Alignment Gap} \\
\midrule
Lifecycle Management 
& $Q_s$ (implicit), $E_s$ (partial) 
& Optimizes resource efficiency and availability; trust state remains opaque to orchestrator 
& RL rewards maximize throughput, not $\mathbf{t}_s$ components;\newline
Intent-to-config mappings lack trust vector decomposition;\newline
Downstream layers must infer $\mathbf{t}_s$ indirectly \\
\addlinespace[0.5em]

Reliability \& SLA Enforcement 
& $Q_s$, $S_s$ (scalar only) 
& Binary or scalar enforcement dominates; continuous vector scoring is rare 
& SLAs are unidimensional, incompatible with WOWA aggregation;\newline
Smart contracts evaluate pass/fail, not continuous trust degrees;\newline
No graduated adaptation mechanism for $\Phi$ propagation \\
\addlinespace[0.5em]

Policy-Driven Governance 
& $S_s$, $F_s$, $P_s$ (boolean) 
& Deterministic constraint enforcement; dynamic trust propagation absent 
& Rules are boolean gates, not continuous variables;\newline
Blockchain lacks $\Delta \mathbf{t}_s$ tracking for feedback;\newline
Formal verification is static, cannot validate runtime trust evolution under changing $\mathbf{n}$ \\
\addlinespace[0.5em]

Runtime Verification 
& All dimensions (unstructured) 
& Rich telemetry generation; structured $\mathbf{t}_s$ feedback is rare 
& Violations not attributed to specific $\mathbf{t}_s$ components;\newline
Audit logs lack real-time $\Psi$ write-back interfaces;\newline
Predictive model outputs are not WOWA-compatible vectors \\
\bottomrule
\end{tabular}
\end{table*}

Table~\ref{tab:service_trisk_gap} synthesizes the cross-layer trust mediation capabilities and TRISK alignment gaps across all four mechanism families. The table reveals a systemic pattern: existing service governance mechanisms excel at local reliability and compliance but consistently lack the vector-valued, propagation-ready trust interfaces required by $\Phi$ and $\Psi$. This gap analysis directly motivates the design principles articulated in Section~\ref{sec:integrated_fabric}.

\subsection{Toward Integrated Trust Mediation in \textsc{TRISK}}

The reviewed service governance mechanisms constitute the trust mediation layer of \textsc{TRISK}, transforming $\mathbf{t}_d$ into executable $\mathbf{t}_s$, exposing runtime evidence for $F_\text{WOWA}$ aggregation, and providing the operational channel for $\Psi$ feedback. Yet significant gaps remain in making services true trust mediators rather than passive compute pipelines. Current mechanisms excel at local reliability and compliance but lack vector-valued, propagation-ready trust interfaces. Scalar SLAs cannot feed WOWA aggregation; boolean policies cannot support graduated $\Phi$ enforcement; unstructured telemetry cannot drive precise $\Psi$ updates.

Bridging these gaps requires redesigning service governance with three principles: (1) \textit{Trust-as-Vector}: exposing $\mathbf{t}_s$ as a five-dimensional, machine-readable attribute rather than scalar SLA; (2) \textit{Propagation-Native Enforcement}: designing policies and contracts that operate on continuous trust degrees, enabling fine-grained trade-offs under $\Phi$; and (3) \textit{Feedback-Structured Telemetry}: annotating runtime violations with trust component attributions and real-time write-back interfaces for $\Psi$. Only through such intentional integration can service orchestration fulfill its role as the dynamic enforcement layer that operationalizes end-to-end trust in Industry 5.0 ecosystems.

\section{Embedding Knowledge as the Semantic Anchor of TRISK} \label{knowledge}

Knowledge governance serves as the semantic anchor that validates whether data-derived facts ($\mathbf{t}_d$) and service-executed actions ($\mathbf{t}_s$) remain consistent with domain rules, physical constraints, and organizational policies. In \textsc{TRISK}, knowledge systems instantiate the trust vector $\mathbf{t}_k = (Q_k, S_k, P_k, F_k, E_k)$ by providing the logical and semantic structures necessary for cross-layer validation. Crucially, they close the trust loop by diagnosing inconsistencies via $\Psi$ and feeding corrective signals back to services and data. This section reviews industrial knowledge governance literature through the \textsc{TRISK} lens, organizing the analysis around four mechanism families by their capability to compute $\mathbf{t}_k$, enforce semantic coherence in $\Phi$, and generate diagnostic feedback for $\Psi$.

\subsection{Characteristics of Industrial Knowledge Systems}
The unique nature of industrial knowledge imposes fundamental constraints on how $\mathbf{t}_k$ can be computed and propagated. We identify three characteristic-driven challenges that existing governance frameworks often fail to address holistically.

\subsubsection{Embedded Domain Expertise vs. Static Semantic Structures}
Industrial knowledge systems encode expertise in ontologies, rules, and knowledge graphs (KGs)~\cite{kg}, capturing relationships among equipment, processes, and operational contexts. However, these representations are typically \textit{static semantic structures} that evolve at design-time scales, while trust propagation via $\Phi$ requires real-time semantic validation. When physical processes drift or new failure modes emerge, static ontologies cannot capture the resulting semantic gaps, causing $Q_k$ and $E_k$ to degrade silently. Knowledge governance must therefore bridge the temporal mismatch between slow-evolving symbolic structures and fast-changing operational realities to prevent trust erosion at the semantic layer.

\subsubsection{Symbolic-Data Convergence vs. Trust Vector Decomposition}
Recent advances integrate symbolic reasoning with data-driven modeling~\cite{sr}: the former provides explicit causal structures and regulatory compliance, while the latter contributes empirical adaptability. Yet this convergence rarely produces a decomposable $\mathbf{t}_k$ vector. Hybrid architectures output scalar confidence scores or binary classifications, not five-dimensional trust attributes suitable for WOWA aggregation (see \eqref{eq_wowa}). A model may achieve high predictive accuracy (high $Q_k$) while violating fairness constraints (low $F_k$), but without vector decomposition, downstream layers cannot perform fine-grained trade-offs. Knowledge governance must therefore expose trust as a multi-dimensional attribute rather than a monolithic score.

\subsubsection{Interpretability Demands vs. Propagation-Opaque Reasoning}
Industrial decisions require transparency, auditability, and compliance~\cite{knowledge,IEC}. Interpretability enables human verification; auditability supports decision path reconstruction; compliance ensures adherence to safety standards. Yet current explainable AI (XAI) mechanisms produce \textit{propagation-opaque reasoning}: explanations are generated for human consumption but not structured as machine-readable $\mathbf{t}_k$ components consumable by $\Psi$. When a knowledge system flags an anomaly, it rarely attributes the violation to specific $S_k$, $P_k$, or $F_k$ degradations with sufficient precision for upstream correction. Knowledge governance must therefore transform interpretability from a human-facing feature into a cross-layer trust signal.

\subsection{Core Mechanisms for Trustworthy Knowledge Governance}

We analyze four mechanism families by their capability to compute $\mathbf{t}_k$, enforce semantic coherence in $\Phi$, and generate diagnostic feedback for $\Psi$. Critically, we assess each against the \textsc{TRISK} requirement of bidirectional trust propagation.

\subsubsection{Knowledge Modeling and Representation: Computing $Q_k$ and $E_k$ with Semantic Precision}
Knowledge modeling forms the structural foundation for $\mathbf{t}_k$ instantiation. Ontologies encode structured expertise, while embedding techniques map symbolic knowledge into vector spaces~\cite{10962554}. Hybrid ontology-embedding models enforce logical constraints during representation learning~\cite{rina}, and alignment-oriented approaches like AutoAlign~\cite{10288249} reconcile heterogeneous KGs without manual schema engineering. Enhanced GNNs~\cite{9681226,10058002} support explainable inference via semantic attention, while SAQE~\cite{11151822} improves complex logical query answering. Multimodal KGs~\cite{10531671} and LLM-KG integration~\cite{10387715} unify textual, visual, and structural evidence, enriching $E_k$. Temporal and hyper-relational embeddings~\cite{10239484} handle evolving process data, providing robustness against knowledge drift.

\textit{TRISK Gap Analysis:} Representation approaches excel at semantic expressiveness but rarely expose $\mathbf{t}_k$ as a machine-readable attribute. Embedding vectors are opaque to symbolic reasoning, making it impossible to verify whether learned representations satisfy specific $F_k$ or $P_k$ constraints. Temporal models track evolution but do not output trust deltas ($\Delta \mathbf{t}_k$) needed for $\Psi$ feedback. LLM-KG integrations improve expressiveness but lack formal guarantees on $S_k$ and $P_k$. Consequently, knowledge models produce semantically rich but trust-opaque representations, forcing downstream layers to infer $\mathbf{t}_k$ indirectly rather than consuming it as explicit metadata. Table~\ref{tab:modeling_comparison} summarizes the trade-offs of representative approaches in interpretability, scalability, and temporal adaptability, yet none natively support five-dimensional trust vector output.

\begin{table*}[htbp]
\centering
\renewcommand{\arraystretch}{0.8}
\footnotesize
\caption{Comparison of Knowledge Representation Approaches}
\label{tab:modeling_comparison}
\begin{tabular}{@{}p{2.8cm} p{2cm} p{1.6cm} p{2.9cm} p{6.4cm}@{}}
\toprule
\textbf{Method} & \textbf{Interpretability} & \textbf{Scalability} & \textbf{Temporal Adaptability} & \textbf{TRISK Limitation} \\
\midrule
Ontology-based~\cite{10962554} & High & Low & Low & Static schema; no runtime $\mathbf{t}_k$ update \\
KG + GNN~\cite{9681226} & Medium & High & Medium & Implicit semantics; no per-dimension trust decomposition \\
LLM-KG~\cite{10387715} & Medium & High & High & No formal $S_k/P_k$ guarantees; prompt-dependent $E_k$ \\
Temporal/Hyper-rel~\cite{10239484} & Medium & Medium & High & Tracks evolution but no $\Delta \mathbf{t}_k$ output for $\Psi$ \\
\bottomrule
\end{tabular}
\end{table*}

\subsubsection{Expert Systems and Rule-Based Reasoning: Enforcing $S_k$, $P_k$, and $F_k$ via Deterministic Logic}
Expert systems operationalize domain expertise into actionable decisions. Modular and mixture-of-experts (MoE) architectures implement distributed reasoning~\cite{9851916,11090031}, enhancing interpretability and robustness. Formal logic (First-Order Logic~\cite{10433728}, decision trees~\cite{10737677}) ensures explainable and policy-compliant inference. Evidential reasoning (BRBs, ER~\cite{11045130}) and Complex Evidential Theory (CET~\cite{9893407}) quantify belief reliability under uncertainty and conflict. These mechanisms directly instantiate $S_k$ (safety compliance), $P_k$ (privacy/regulatory adherence), and $F_k$ (fairness via explicit constraint encoding).

\textit{TRISK Gap Analysis:} Rule-based systems provide deterministic guarantees but lack continuous trust degrees required for WOWA aggregation. Boolean rule satisfaction (pass/fail) cannot express graduated $S_k$ or $F_k$ degradation. Evidential reasoning outputs belief/plausibility pairs but not five-dimensional $\mathbf{t}_k$ vectors. MoE gating weights optimize accuracy, not trust component balance. Consequently, expert systems act as rigid validators rather than adaptive trust mediators, unable to support the fine-grained, context-responsive enforcement required by $\Phi$ under partial trust degradation.

\subsubsection{Semantic Alignment and Interoperability: Preserving $Q_k$ and $E_k$ Across Boundaries}
Semantic alignment ensures heterogeneous knowledge across domains and modalities can be coherently integrated. Entity/relation alignment frameworks merge distributed KGs~\cite{9954199}, while multimodal integration links textual, visual, and sensor-derived information~\cite{11054304}. MAMN~\cite{9723597} and IISAN-Versa~\cite{11153882} unify heterogeneous modality spaces via adaptive dimension-alignment. Domain-adaptive frameworks (CAF~\cite{9803869}) reduce domain shifts while preserving semantic fidelity. Hierarchical relation clustering and semantic smoothing (BioKDN~\cite{10216353}) improve cross-domain reasoning and reduce inconsistency~\cite{10706014}. These mechanisms preserve $Q_k$ (semantic accuracy) and $E_k$ (explainability) across trust boundaries.

\textit{TRISK Gap Analysis:} Alignment methods optimize for structural or semantic similarity metrics (H@k, MRR) but do not verify trust vector consistency across boundaries. Two aligned entities may have identical embeddings but divergent $P_k$ or $F_k$ values due to differing regulatory contexts. Cross-domain transfer preserves semantics but not trust compositionality: $\mathbf{t}_k$ from the source domain cannot be directly aggregated with target domain $\mathbf{t}_k$ without re-normalization. Consequently, alignment enables syntactic interoperability but not trust-propagatable semantic coherence, leaving $\Phi$ vulnerable to hidden trust discontinuities at domain interfaces.

\subsubsection{Continuous Validation and Updating: Generating $\Psi$ Feedback via Adaptive $Q_k$ and $S_k$}
Continuous validation maintains knowledge reliability under evolving conditions. Incremental update frameworks integrate new data without full retraining~\cite{9834133}. Confidence-based optimization~\cite{11045130} refines knowledge via uncertainty estimation. AEKE~\cite{10239484} dynamically adjusts aggregation weights based on triple confidence. Adaptive mechanisms address concept drift~\cite{10103675,9891841,wang2020multiscale}, while evaluation-driven feedback loops trigger controlled updates~\cite{11048971}. Adversarial testing frameworks (ShillingREC~\cite{11045936}, LOKI~\cite{9806383}) validate system defenses, driving resilient model evolution. These mechanisms sustain $Q_k$ and $S_k$ over time and generate evidence for $\Psi$.

\textit{TRISK Gap Analysis:} Validation mechanisms detect degradation but rarely structure feedback as $\mathbf{t}_k$ components consumable by $\Psi$. Drift detectors signal performance drop but do not attribute it to specific $Q_k$, $S_k$, or $F_k$ violations. Confidence scores are scalar, not vector-valued. Adversarial tests identify vulnerabilities but do not output trust deltas ($\Delta \mathbf{t}_k$) for upstream correction. Feedback loops are typically unidirectional (model $\to$ data) rather than bidirectional (knowledge $\to$ service $\to$ data). Consequently, validation generates rich but unstructured diagnostic signals, forcing upstream layers to perform costly inference rather than consuming pre-computed $\mathbf{t}_k$ feedback for precise trust repair.

Table~\ref{tab:knowledge_trisk_gap} synthesizes the cross-layer trust capabilities and TRISK alignment gaps across all four mechanism families. The table reveals a systemic pattern: existing knowledge governance mechanisms excel at semantic expressiveness and local validation but consistently lack vector-valued, propagation-ready trust interfaces required by $\Phi$ and $\Psi$. This gap analysis directly motivates the design principles articulated in Section~\ref{sec:integrated_fabric}.

\begin{table*}[htbp]
\centering
\renewcommand{\arraystretch}{1.1}
\footnotesize
\caption{Cross-Layer Trust Capability of Knowledge Governance Mechanisms}
\label{tab:knowledge_trisk_gap}
\begin{tabular}{@{}p{2.6cm} p{3.0cm} p{3.8cm} p{6.0cm}@{}}
\toprule
\textbf{Mechanism Family} & \textbf{$\mathbf{t}_k$ Components Supported} & \textbf{Propagation Strength} & \textbf{TRISK Alignment Gap} \\
\midrule
Knowledge Modeling \& Representation 
& $Q_k$, $E_k$ (implicit) 
& Semantically expressive; trust state opaque to orchestrator 
& Embeddings lack per-dimension trust decomposition;\newline
Temporal models track evolution but no $\Delta \mathbf{t}_k$ output;\newline
LLM-KG lacks formal $S_k/P_k$ guarantees \\
\addlinespace[0.5em]

Expert Systems \& Rule-Based Reasoning 
& $S_k$, $P_k$, $F_k$ (boolean) 
& Deterministic constraint enforcement; continuous trust degrees absent 
& Boolean satisfaction incompatible with WOWA;\newline
Evidential outputs not five-dimensional;\newline
MoE gating optimizes accuracy, not trust balance \\
\addlinespace[0.5em]

Semantic Alignment \& Interoperability 
& $Q_k$, $E_k$ (cross-boundary) 
& Structural/semantic alignment; trust vector consistency unverified 
& Aligned entities may have divergent $P_k/F_k$;\newline
No trust compositionality across domains;\newline
Syntactic interoperability $\neq$ trust-propagatable coherence \\
\addlinespace[0.5em]

Continuous Validation \& Updating 
& $Q_k$, $S_k$ (diagnostic) 
& Degradation detection; structured $\mathbf{t}_k$ feedback rare 
& Drift signals not attributed to specific $\mathbf{t}_k$ components;\newline
Confidence scores scalar, not vector-valued;\newline
Feedback loops unidirectional, lacking $\Psi$ write-back \\
\bottomrule
\end{tabular}
\end{table*}

\subsection{Knowledge Governance as the Trust Engine of \textsc{Trisk}}

In \textsc{TRISK}, knowledge governance is the semantic validator and trust amplifier that closes the trust loop across layers. It operationalizes $E_k$, $F_k$, and semantic consistency by evaluating whether service decisions and data transformations comply with domain rules, physical laws, and business policies. When inconsistencies arise, knowledge governance downgrades $\mathbf{t}_k$, triggers corrective actions in service orchestration via $\Psi$, or calls for data reinterpretation and lineage revision, preventing silent propagation of erroneous or biased decisions. Conversely, trusted data streams ($\mathbf{t}_d$) continuously update knowledge models, while trusted services ($\mathbf{t}_s$) operationalize knowledge into verifiable actions.

Yet significant gaps remain in making knowledge systems true trust engines rather than passive semantic repositories. Current mechanisms excel at local expressiveness and validation but lack vector-valued, propagation-ready trust interfaces. Static ontologies cannot capture real-time semantic drift; boolean rules cannot support graduated $\Phi$ enforcement; alignment metrics do not verify trust consistency; validation signals are unstructured for $\Psi$ consumption. Bridging these gaps requires redesigning knowledge governance with three principles: (1) \textit{Trust-as-Vector}: exposing $\mathbf{t}_k$ as a five-dimensional, machine-readable attribute; (2) \textit{Propagation-Native Semantics}: designing representations and rules that operate on continuous trust degrees, enabling fine-grained trade-offs under $\Phi$; and (3) \textit{Feedback-Structured Diagnostics}: annotating validation signals with trust component attributions and real-time write-back interfaces for $\Psi$. Only through such intentional integration can knowledge governance fulfill its role as the semantic anchor that sustains end-to-end trust in Industry 5.0 ecosystems.

\section{Industrial Implementation and Cross-Industry Implications} 
\label{sec:implementation}

To validate the practical efficacy of \textsc{TRISK} in addressing governance fragmentation, this section examines its deployment within a large-scale 3C discrete manufacturing enterprise. We first detail the internal coordination mechanism that binds data, service, and knowledge governance through hardware-software co-assurance, then demonstrate how this mechanism resolves specific industrial pain points via trust vector propagation, and finally formalize the framework's adaptability across different industrial contexts through parameterized weight configuration.


\subsection{Internal Coordination Mechanism: Unifying the Governance Core}
\label{sec:internal_mech}
\begin{figure}[htbp]
    \centering 
    \includegraphics[width=\linewidth]{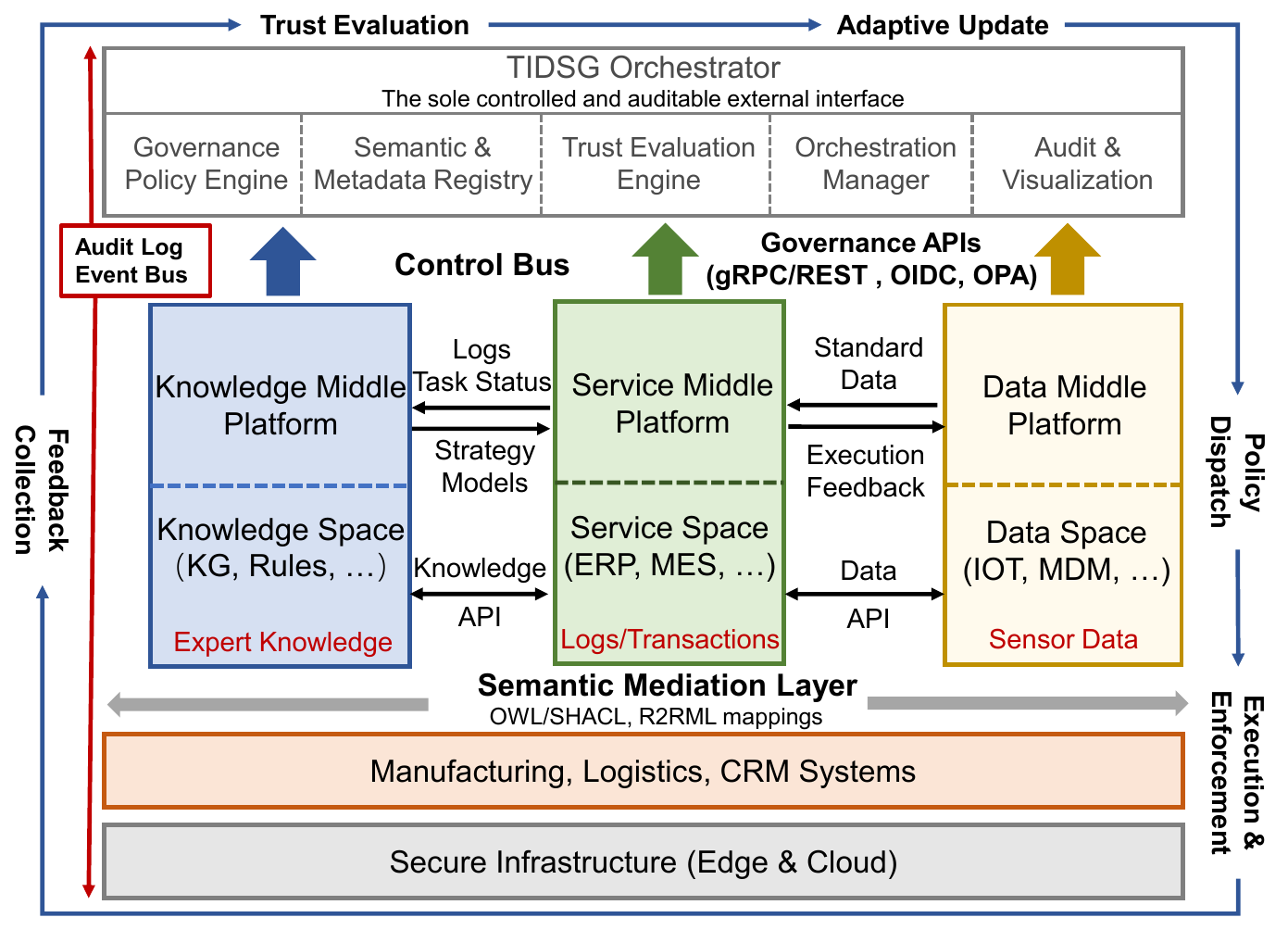} 
    \caption{Internal Coordination Mechanism of \textsc{TRISK}. The orchestration layer unifies data, service, and knowledge governance atop a hardware-software co-assurance foundation. Trust vectors ($\mathbf{t}_d, \mathbf{t}_s, \mathbf{t}_k$) propagate bidirectionally via $\Phi$ and $\Psi$.}
    \label{fig:Inter} 
\end{figure}

The primary cause of governance fragmentation lies in the misalignment between data producers, service executors, and decision-makers. \textsc{TRISK} resolves this by establishing a centralized \textit{Orchestration Layer} that acts as the coordination brain (Fig.~\ref{fig:Inter}). This layer does not merely route data but actively manages the lifecycle of trust vectors ($\mathbf{t}_d, \mathbf{t}_s, \mathbf{t}_k$) across three interconnected platforms, anchored by a hardware-software co-assurance foundation.

\subsubsection{Three Governance Platforms}
\begin{itemize} 
    \item \textit{Data Platform:} Aggregates heterogeneous streams (IoT sensors, MES logs, quality inspection) and computes $\mathbf{t}_d$ via semantic validation and lineage tracking. It ensures raw data meets quality preconditions required by downstream $\Phi$ propagation.
    \item \textit{Service Platform:} Translates business logic (ERP, SCM, APS) into executable workflows and computes $\mathbf{t}_s$. It enforces policy compliance and exposes runtime evidence for $\Psi$ feedback generation.
    \item \textit{Knowledge Platform:} Hosts domain ontologies, rule bases, and analytical models to compute $\mathbf{t}_k$. It provides reasoning capability to diagnose systemic inconsistencies and refine governance policies via $\Psi$.
\end{itemize} 

These platforms are harmonized through a \textit{Semantic Mediation Layer} using OWL and SHACL standards, ensuring that a defect code from a machine sensor is interpreted consistently by quality management services and predictive maintenance models. This semantic consistency is the cornerstone of resolving fragmentation.

\subsubsection{Hardware-Software Co-Assurance as Trust Anchor}
Crucially, the orchestration layer operates atop a hardware-software co-assurance foundation that anchors $S_s$ and $E_s$ in verifiable physical and logical guarantees. On the hardware side, Trusted Execution Environments (TEEs)~\cite{TEE} provide tamper-proof computing zones for trust evaluation and policy enforcement, ensuring that $\mathbf{t}_s$ computations cannot be subverted by compromised OS or middleware. Hardware-rooted attestation enables remote verification of service integrity before trust propagation. On the software side, policy-driven APIs, encryption protocols (TLS 1.3), and authentication frameworks (gRPC~\cite{gRPC}, OIDC~\cite{OIDC}) enforce controlled communication among components. Audit logs are cryptographically chained to TEE-sealed timestamps, providing immutable evidence for $E_s$ and enabling compliant $\Psi$ feedback. This co-assurance transforms trust from a purely logical construct into a physically grounded, auditable capability.

\subsection{Case Study: Resolving Fragmentation in 3C Manufacturing}
\label{sec:case_study} 

The enterprise operates multiple plants involving precision machining, automated assembly, and logistics. Prior to \textsc{TRISK}, the ecosystem was fragmented: machine data used proprietary protocols, MES/ERP systems had conflicting scheduling logic, and expert knowledge was trapped in technicians' manuals. This resulted in frequent production halts due to untraceable data discrepancies.

\subsubsection{The Challenge of Semantic Fragmentation}
A recurring issue highlighted the core problem: Surface-mount technology (SMT) machines occasionally stopped due to material errors. Data logs ($\mathbf{t}_d$) indicated a feeder error, MES ($\mathbf{t}_s$) recorded a material shortage, and maintenance teams ($\mathbf{t}_k$) suspected static electricity damage. Without unified governance, these were treated as isolated incidents with incompatible trust interpretations, leading to wasted downtime while teams finger-pointed. The root cause was \textit{trust vector misalignment}: each layer computed its local trust state independently, with no $\Phi$-mediated reconciliation or $\Psi$-driven correction.

\subsubsection{TRISK Intervention: Trust Vector Propagation and Feedback}
\begin{figure}[htbp]
    \centering
    \includegraphics[width=\linewidth]{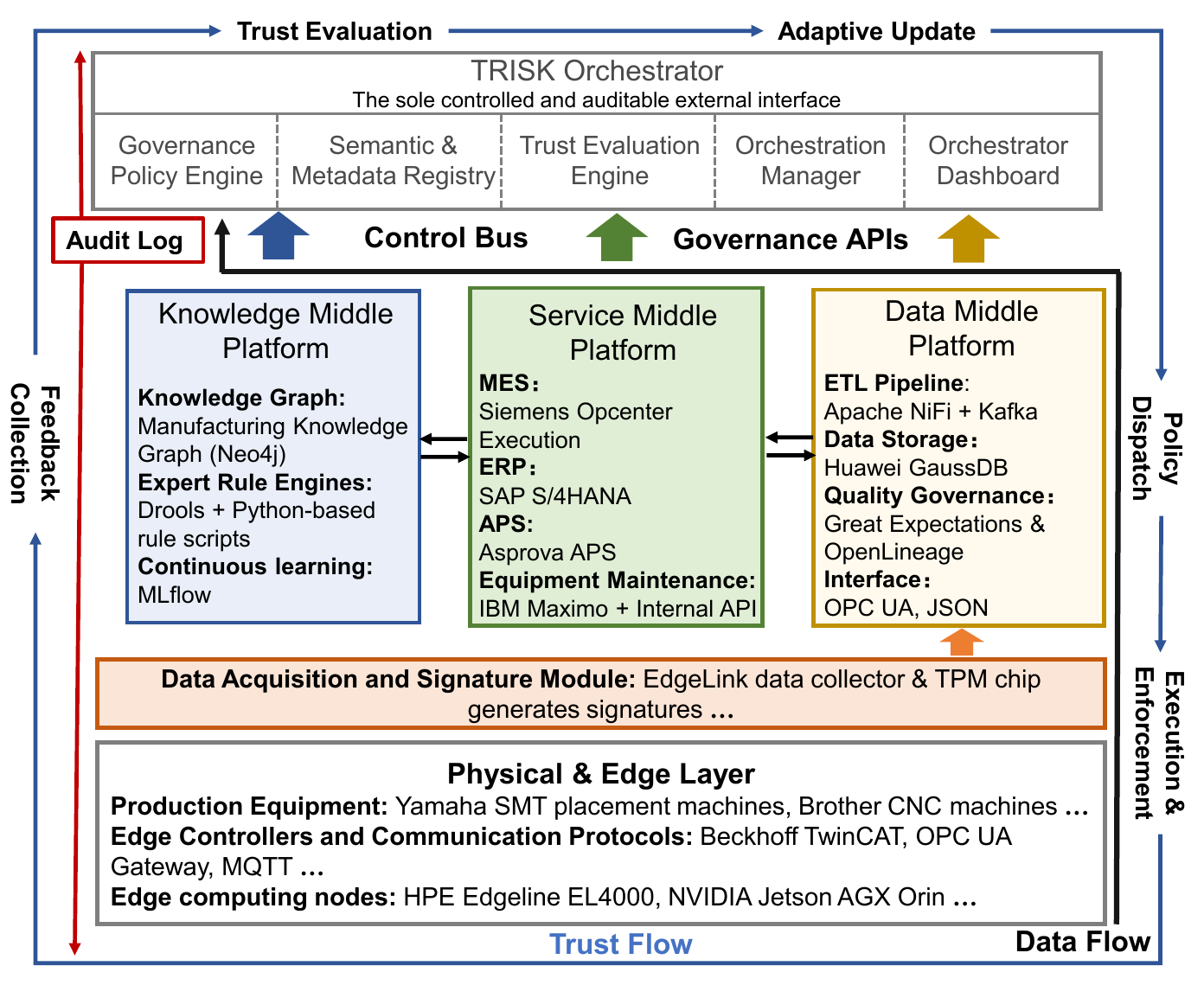}
    \caption{System architecture of \textsc{TRISK} deployment in 3C manufacturing. Upward data flow carries validated $\mathbf{t}_d$; downward trust flow propagates $\Psi$ feedback as updated policies and verification results.}
    \label{arch}
\end{figure}

\textsc{TRISK} restructured operations by binding layers through explicit trust vector propagation (Fig.~\ref{arch}):

\begin{itemize}
    \item \textit{Semantic Alignment via $\Phi$:} The Semantic Mediation Layer mapped disparate error codes to a unified ontology. It established that the SMT feeder error ($Q_d$ degradation) was physically correlated with a specific vibration frequency. This allowed $\Phi$ to reject the MES material shortage interpretation (which assumed high $Q_d$) and instead propagate a corrected $\mathbf{t}_d$ with degraded $Q_d$ but elevated $E_d$ (vibration evidence) to the service layer.
    
    \item \textit{Service Orchestration via $\mathbf{t}_s$ Adaptation:} Upon receiving the corrected $\mathbf{t}_d$, the Service Platform computed $\mathbf{t}_s$ with degraded $Q_s$ (due to uncertain input) but elevated $S_s$ (safety-first response). It automatically halted the assembly line via PLC API and locked the affected material batch, invoking a deep cleaning service rather than generic replenishment. This graduated adaptation, enabled by continuous trust degrees rather than binary SLAs, directly linked data anomaly to correct corrective action.
    
    \item \textit{Knowledge Feedback via $\Psi$:} The incident was logged with full $\mathbf{t}_d, \mathbf{t}_s, \mathbf{t}_k$ attribution. $\Psi$ diagnosed that this vibration pattern should be classified as a maintenance-required event ($S_k$ update) rather than a production error ($Q_k$ preservation). This updated rule was pushed back to data validation policies (adjusting $Q_d$ computation) and service orchestration logic (adjusting $\mathbf{t}_s$ adaptation thresholds), closing the trust loop.
\end{itemize}

\subsubsection{Observed Effects}
By resolving semantic and procedural fragmentation through trust vector propagation, TRISK changed the way operational evidence was collected, interpreted, and audited in the manufacturing workflow. Instead of relying on isolated alarms or manual cross-checking across MES, equipment logs, and technician knowledge, integrated $\mathbf{t}_d$-$\mathbf{t}_s$-$\mathbf{t}_k$ traces provided a unified view of data anomalies, service actions, and knowledge-rule updates. This helped maintenance engineers distinguish material-related, equipment-related, and rule-related causes more systematically. The $\Phi$-mediated semantic alignment also reduced ambiguity between data records and service decisions. In the SMT case, inconsistent interpretations of feeder errors, material shortages, and maintenance rules were reconciled through a shared ontology and trust-aware service adaptation. As a result, corrective actions became more targeted, such as locking affected batches or invoking maintenance-specific workflows rather than triggering generic replenishment procedures.

Furthermore, TEE-sealed logs and trust-attributed execution records strengthened auditability. Each intervention could be traced back to the corresponding data evidence, service decision, knowledge rule, and policy update, providing verifiable support for post-event review and compliance inspection. These observations suggest that hardware-software co-assurance can improve operational transparency and reduce governance friction, although large-scale quantitative validation remains future work.

\subsection{Cross-Industry Applicability and Adaptability}
\label{sec:cross_industry} 

While the 3C case validates \textsc{TRISK}'s core mechanics, the framework is designed for cross-industry adaptability through parameterized configuration of WOWA weight vectors $\mathbf{w}$ and feedback timescales $\tau_\Psi$. The modular governance core allows domain-specific adjustments without architectural changes.

\begin{itemize}
    \item \textit{Energy Sector (Smart Grid):} Primary challenge shifts from assembly precision to real-time risk. WOWA weights prioritize $S_s$ and $Q_s$ over $F_s$ ($w_{S_s}, w_{Q_s} \gg w_{F_s}$). Semantic mediation harmonizes IEC 61850 device protocols with market trading logic. Feedback timescale $\tau_\Psi$ is milliseconds-seconds, requiring FPGA-accelerated runtime verification for grid stabilization.
    
    \item \textit{Healthcare/Pharma Supply Chain:} Privacy ($P$) and provenance ($E$) dominate. WOWA weights prioritize $P_d, P_s, P_k$ and $E_d, E_s, E_k$ ($w_P, w_E \gg w_Q$). Knowledge platform enforces HL7 FHIR regulatory ontologies. Feedback timescale $\tau_\Psi$ is weeks-months (clinical outcomes), emphasizing long-term audit trails and immutable lineage for FDA 21 CFR Part 11 compliance over real-time adaptation.
    
    \item \textit{Automotive Manufacturing:} Safety ($S$) and fairness ($F$) in autonomous testing require balanced weights ($w_S \approx w_F$). Feedback integrates both real-time telemetry ($\tau_\Psi \sim$ seconds) and long-term field failure data ($\tau_\Psi \sim$ months), necessitating multi-timescale $\Psi$ aggregation.
\end{itemize}

Table~\ref{tab:cross_industry_adaptation} summarizes the parameterized adaptation across industries, demonstrating that \textsc{TRISK} is not a rigid stack but a flexible governance paradigm whose trust dimensions can be formally reconfigured to address unique manifestations of fragmentation.

\begin{table*}[htbp]
\centering
\renewcommand{\arraystretch}{1.1}
\footnotesize
\caption{Parameterized Cross-Industry Adaptation of \textsc{TRISK}}
\label{tab:cross_industry_adaptation}
\begin{tabular}{@{}p{2.2cm} p{3.0cm} p{3.5cm} p{3.5cm} p{3.0cm}@{}}
\toprule
\textbf{Industry} & \textbf{Dominant Trust Dimensions} & \textbf{WOWA Weight Priority} & \textbf{Feedback Timescale ($\tau_\Psi$)} & \textbf{Key Enforcement Mechanism} \\
\midrule
3C Manufacturing & Quality ($Q$), Explainability ($E$) & $w_Q, w_E > w_S > w_F, w_P$ & Minutes-Hours (production cycle) & Semantic mediation + TEE-sealed audit \\
\addlinespace[0.3em]
Energy / Smart Grid & Safety ($S$), Quality ($Q$) & $w_S, w_Q \gg w_F, w_P$ & Milliseconds-Seconds (grid stability) & FPGA runtime verification + IEC 61850 \\
\addlinespace[0.3em]
Healthcare / Pharma & Privacy ($P$), Provenance ($E$) & $w_P, w_E \gg w_Q, w_F$ & Weeks-Months (clinical outcomes) & Immutable lineage + HL7 FHIR ontology \\
\addlinespace[0.3em]
Automotive & Safety ($S$), Fairness ($F$) & $w_S \approx w_F > w_Q$ & Multi-scale (telemetry + field data) & Multi-timescale $\Psi$ aggregation \\
\bottomrule
\end{tabular}
\end{table*}

\begin{figure}[htbp]
    \centering 
    \includegraphics[width=0.75\linewidth]{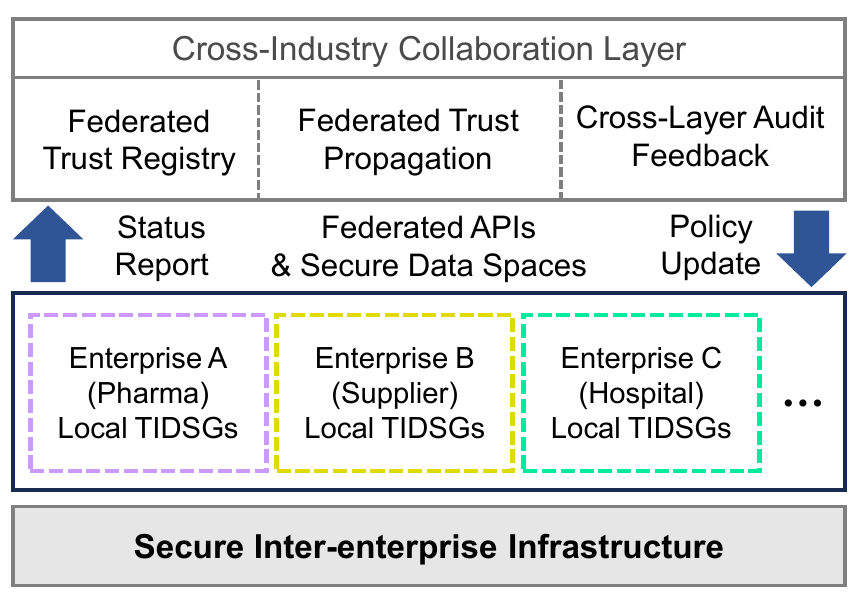} 
    \caption{Extension to Cross-Industry Collaboration. Federated \textsc{TRISK} nodes enable value-chain trust propagation via standardized $\mathbf{t}$-vector exchange while preserving data sovereignty through TEE-attested secure enclaves.}
    \label{fig:Outer} 
\end{figure}

Beyond single-enterprise deployment, \textsc{TRISK} extends to cross-industry collaboration (Fig.~\ref{fig:Outer}). Multiple stakeholders connect local \textsc{TRISK} systems through federated APIs and secure data spaces. Each enterprise maintains internal governance autonomy but contributes standardized trust reports ($\mathbf{t}_d, \mathbf{t}_s, \mathbf{t}_k$ summaries) to a federated trust registry. TEE-attested enclaves enable policy alignment and federated learning without sharing raw production data. For example, a 3C manufacturer can share process capability insights ($Q_k$ aggregates) with component suppliers to optimize design tolerances, or integrate field repair data ($E_s$ feedback) from service centers to refine upstream assembly parameters. Through federated trust propagation, \textsc{TRISK} evolves from an enterprise framework into infrastructure for collaborative industrial ecosystems, enabling scalable, explainable, and secure cooperation while preserving data sovereignty.

\section{Challenges and Future Research Agenda} \label{future}
The \textsc{TRISK} framework establishes a foundational paradigm for trustworthy industrial governance; however, bridging the gap between theoretical formulation and industrial practice requires overcoming significant interdisciplinary hurdles. This section analyzes the challenges across theoretical, methodological, and ecosystem dimensions, and outlines a strategic roadmap to advance the state of the art.

\subsection{Theoretical and Conceptual Challenges}
\label{sec:theory_challenges}

While Section~\ref{found} proposes a formalized representation of the \textsc{TRISK} loop, translating this conceptual model into a universally accepted theoretical framework faces significant hurdles. The current lack of unified theoretical foundations hinders the reproducibility and verifiability of cross-layer governance. Three specific conceptual gaps require urgent attention.

\paragraph{Standardization and Acceptance of Cross-Layer Trust Semantics}
Although we have defined layer-specific trust states ($T_i$) and a cross-layer propagation operator ($\mathcal{F}$) in Section~\ref{found}, the broader industrial and academic community lacks consensus on the formal semantics of trust. Current implementations rely on disparate empirical indicators (e.g., data accuracy, SLA compliance) that are rarely interoperable. Future work must focus on standardizing these semantics, combining formal logic, information theory, and probabilistic reasoning, to ensure that the mathematical constructs proposed in \textsc{TRISK} are not only theoretically sound but also practically measurable and comparable across heterogeneous systems.

\paragraph{Characterizing Intrinsic Trade-offs in Trust Optimization}
Industrial governance often involves navigating conflicting objectives, such as the tension between privacy preservation and data utility, or between model explainability and real-time performance. While our model in Section~\ref{found} aggregates Quality, Security, Privacy, Fairness, and Explainability (QSPFE), a rigorous theoretical framework is still needed to characterize the dynamic trade-offs among these dimensions. Developing a multi-objective optimization framework that can adaptively balance these constraints based on context-specific priorities (e.g., safety-critical vs. efficiency-driven scenarios) remains a core theoretical challenge.

\paragraph{Modeling the Dynamics of Trust Propagation and Degradation}
A critical gap lies in the dynamic modeling of trust throughout the full governance lifecycle. As highlighted in Section~\ref{found}, trust is not static; it evolves and propagates through the data, service, and knowledge layers. However, existing studies seldom model how trust degrades due to latency, interference, or adversarial attacks in complex cyber-physical systems. Bridging this gap requires extending our current model to rigorously capture the bidirectional propagation and correction processes, transforming \textsc{TRISK} from a static model into a dynamic, self-correcting theoretical paradigm.

\begin{figure*}[t] 
    \centering 
    \includegraphics[width=0.9\linewidth]{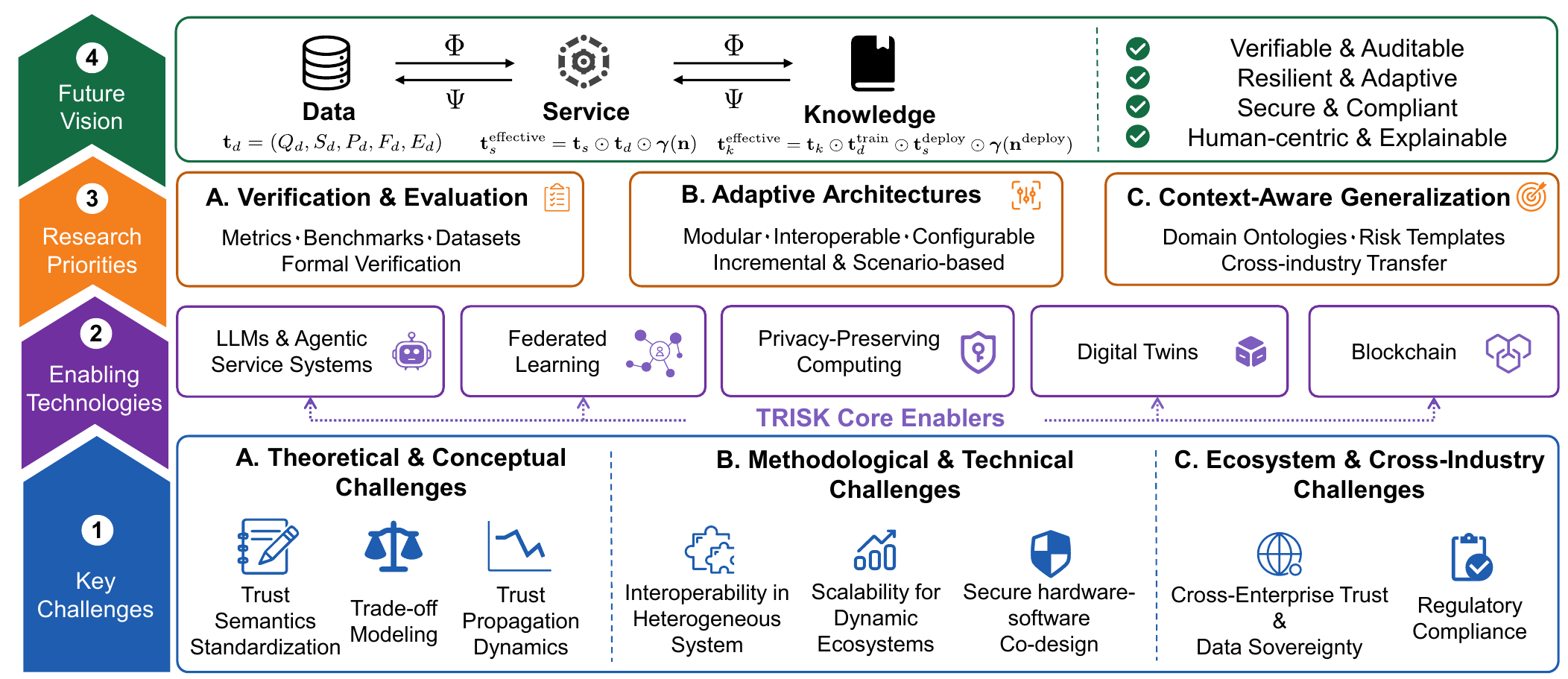} 
    \caption{Future research roadmap for \textsc{TRISK}. Building upon the trust propagation and feedback mechanisms of the \textsc{TRISK} framework, the roadmap highlights key research challenges, emerging technology enablers, strategic priorities, and future directions for trustworthy industrial intelligence.} 
    \label{futuremap} 
\end{figure*}

\subsection{Methodological and Technical Challenges}
\label{sec:method_challenges}

Translating the \textsc{TRISK} paradigm into operational systems demands resolving deep-seated methodological conflicts between legacy industrial infrastructure and modern governance requirements. These challenges represent significant barriers to scalability and interoperability.

\paragraph{Achieving Interoperability in Heterogeneous Environments}
Industrial enterprises are characterized by complex, heterogeneous landscapes comprising legacy systems (ERP, MES, PLCs) and modern IoT platforms. These systems operate on divergent protocols and data models, creating ``governance silos''. Enforcing unified policies requires more than standardization; it necessitates advanced semantic mediation techniques capable of dynamically aligning governance logic across distributed, multi-vendor environments.

\paragraph{Scaling Governance for Dynamic Ecosystems}
The expansion of digital supply chains necessitates governance mechanisms that scale to unprecedented volumes of data and services. Traditional centralized monitoring architectures are ill-suited for real-time trust assessment in such dynamic ecosystems. Future research must explore decentralized orchestration and federated trust propagation to handle these workloads, while simultaneously addressing the computational overhead and integration costs that currently hinder industrial deployment.

\paragraph{Integrating Secure Hardware-Software Co-Design}
Extending trust assurance to the network edge requires a tight coupling of physical hardware and software policies. While Trusted Execution Environments (TEEs) and secure enclaves offer verifiable computation, integrating them with high-level policy engines remains difficult. The challenge lies in designing lightweight, reconfigurable governance modules that can operate within the stringent resource constraints of industrial edge devices without compromising security guarantees.

\subsection{Ecosystem and Cross-Industry Challenges}
\label{sec:ecosystem_challenges}

The extension of \textsc{TRISK} beyond the enterprise boundary introduces complex socio-technical challenges related to trust boundaries, regulation, and accountability.

\paragraph{Cross-Enterprise Trust and Data Sovereignty}
In global supply networks, governance must function across organizational boundaries where trust is not inherent. Ensuring trust continuity while respecting data sovereignty and ownership rights requires novel federated governance mechanisms. These mechanisms must enable collaborative optimization (e.g., joint quality control) without exposing sensitive raw data, thereby aligning business collaboration with strict confidentiality requirements.

\paragraph{Harmonizing Standards and Regulatory Compliance}
Different industrial sectors, ranging from aerospace to pharmaceuticals, operate under vastly different regulatory regimes and risk models. A one-size-fits-all governance approach is infeasible. Future work must focus on developing a domain-neutral yet context-aware governance paradigm that can dynamically adapt to specific compliance templates and safety requirements, facilitating cross-industry adoption without sacrificing regulatory rigor.

\subsection{Opportunities with Emerging Technologies}
\label{sec:emerging_tech}

The resolution of the aforementioned challenges is closely tied to the synergistic integration of \textsc{TRISK} with several cutting-edge technological paradigms.

\paragraph{LLMs and Agentic Service Systems for Cognitive Orchestration}
Large Language Models (LLMs) and autonomous agents offer a pathway to endow \textsc{TRISK} with cognitive capabilities. By transforming unstructured industrial documents (SOPs, manuals) into machine-interpretable knowledge, LLMs can enhance the reasoning capacity of the knowledge layer. Furthermore, agentic service systems can enable multi-agent collaboration, where autonomous agents representing different governance roles (e.g., Auditor, Compliant Checker) collectively maintain trust continuity through interactive reasoning.

\paragraph{Federated Learning and Privacy-Preserving Computation}
To address the tension between data utility and privacy, federated learning (FL) and privacy-enhancing technologies (PETs) such as secure multi-party computation (SMPC) are pivotal. These technologies align perfectly with the federated architecture of \textsc{TRISK}, allowing industrial consortia to collaboratively train predictive models on distributed data assets. This ensures that sensitive production data remain localized while still contributing to global trust evaluation and optimization.

\paragraph{Blockchain and Digital Twins as Verification Infrastructures}
Blockchain technology provides an immutable ledger for recording data provenance and audit trails, ensuring non-repudiation in inter-organizational workflows. Complementarily, Digital Twins serve as governance sandboxes. By mirroring physical production systems, digital twins allow \textsc{TRISK} policies and AI-driven decisions to be simulated, stress-tested, and optimized in a risk-free environment before real-world deployment, thereby closing the loop between digital governance and physical operations.

\subsection{A Roadmap for Future Research}
\label{sec:roadraph}

To transform \textsc{TRISK} into a scalable and intelligent governance infrastructure, future research should focus on the following strategic directions, as illustrated in Fig.~\ref{futuremap}.

\begin{enumerate}
    \item \textit{Establishing Verification and Evaluation Frameworks:} Develop standardized benchmarks that combine quantitative metrics (e.g., lineage consistency, model interpretability scores) with qualitative assessments. This includes creating public datasets and simulation environments to facilitate reproducible comparative studies on trustworthy governance.
    
    \item \textit{Designing Adaptive Reference Architectures:} Move beyond conceptual models to define concrete architectural blueprints. These should incorporate modular, hardware-assisted trust modules and standardized semantic interfaces, enabling enterprises to deploy \textsc{TRISK} incrementally without disrupting existing legacy operations.
    
    \item \textit{Realizing Context-Aware Generalization:} Explore methodologies that allow \textsc{TRISK} to adapt its core principles to domain-specific constraints. By incorporating industry-specific ontologies and risk templates, the framework can evolve into a truly context-aware governance paradigm capable of supporting the digital transformation of diverse sectors.
\end{enumerate}

\section{Conclusion}
\label{conclusion}

This paper presents \textsc{TRISK} (TRusted Industrial Data-Service-Knowledge governance), a conceptual and taxonomic framework for trustworthy industrial intelligence. By integrating data governance, service governance, and knowledge governance under a unified trust-centered perspective, \textsc{TRISK} addresses the fragmentation that currently limits end-to-end reliability, accountability, compliance, and explainability in industrial systems. We first establish the theoretical foundations of multidimensional trust, including quality, security, privacy, fairness, and explainability, and formalize how trust is constructed, propagated, aggregated, and fed back across data, service, and knowledge layers. Building on this foundation, we synthesize representative studies to show how data governance constructs the initial trust state, service governance mediates trustworthy execution, and knowledge governance provides semantic validation and feedback adaptation.

We further discuss how \textsc{TRISK} can be operationalized through secure foundations, semantic mediation, governance APIs, trust evaluation, and orchestration mechanisms. The manufacturing example shows how \textsc{TRISK} transforms fragmented data, service, and knowledge operations into an adaptive governance system with traceable trust flows and closed-loop correction. Beyond single enterprises, it also supports federated and cross-industry governance across organizational and regulatory boundaries. \textsc{TRISK} marks a shift from data-centric management to trust-centric industrial intelligence, offering a pathway toward adaptive, verifiable, human-aligned, and sustainable governance for Industry 5.0.

\bibliographystyle{IEEEtran}
\bibliography{ref}


 





\end{document}